\documentclass{sig-alternate} 
\usepackage{mathptmx} 

\newcommand{\ignore}[1]{}
\usepackage{fancyhdr}
\usepackage[normalem]{ulem}
\usepackage[hyphens]{url}
\usepackage{microtype}
\usepackage[font=footnotesize]{caption}
\usepackage{subcaption}
\usepackage{verbatim}
\usepackage{color,soul}

\usepackage{amsmath}
\usepackage{algorithm}
\usepackage[noend]{algpseudocode}
\usepackage[bookmarks=true,breaklinks=true,letterpaper=true,colorlinks,linkcolor=black,citecolor=blue,urlcolor=black]{hyperref}
\usepackage{multirow}

\usepackage{listings}
\usepackage{courier}
\usepackage{tikz}
\usepackage{xcolor}
\newcommand*\circled[1]{\tikz[baseline=(char.base)]{
    \node[shape=circle,fill,inner sep=1pt] (char) {\textcolor{white}{#1}};}}
\usepackage{fixltx2e}

\fancypagestyle{firstpage}{
  \fancyhf{}

  \fancyfoot[C]{\thepage}
}  

\title{Remote Control: A Simple Deadlock Avoidance Scheme for Modular System on Chip} 
\author{Pritam Majumder, Sungkeun Kim, Jiayi Huang, Ki Hwan Yum, Eun Jung Kim \\Texas A\&M University
\\(pritam2309, ksungkeun84, jyhuang, yum, ejkim)@tamu.edu}

\begin{document}
\maketitle
\thispagestyle{firstpage}
\pagestyle{plain}

\begin{abstract}
\label{abstract}
The increase in design cost and complexity have motivated designers to adopt modular design of System on Chip (SoC) by integrating independently designed small chiplets. However, it introduces new challenges for correctness validation, increasing chances of forming deadlock in the system involving multiple chiplets. Although there have been many solutions available for deadlock freedom in flat networks, the study on deadlock issue in chiplet-based systems is still in its infancy. A recent study suggests adding extra turn restrictions as a viable solution for this problem. However, imposing extra turn restrictions reduces chiplet design flexibility and interposer design complexity. In addition, it may lead to non-minimal route and traffic imbalance forming hotspots, resulting in high latency and low throughput.

We propose \textit{Remote Control (RC)}, a simple routing oblivious deadlock avoidance scheme. Our proposal is based on two key observations. First, packets with destinations in the current chiplet are blocked by packets with destinations outside the chiplet. Second, deadlock always involves multiple boundary routers\footnote{Boundary routers are the edge routers in the chiplet.}. Hence, we segregate different traffics to alleviate mutual blocking at the chiplet's boundary routers. Along with guarantee of deadlock freedom and performance enhancements, our simple \textit{RC} scheme also provides more routing flexibility to both chiplet and SoC designers, as compared to the state-of-the-art.
\end{abstract}

\section{Introduction}
\label{intro}
With the advancements in silicon technology, Systems-on-Chip (SoCs) are becoming more complex and expensive, which motivates the designers to break the whole SoC into multiple small independent {\em chiplets} for reducing design cost and achieve better scalability. Modular design of SoCs using 2.5D integration technology is a total paradigm shift from  monolithic SoC design to hierarchical SoC design~\cite{green16common,demir2014galaxy,cianchetti2010implementing,kannan2015enabling,iyer2016heterogeneous}. 
It allows the architects to design smaller independent chiplets (e.g., CPU, GPU, and accelerators) with low cost and complexity, and integrate them together on an interposer, creating heterogeneous chiplet-based architectures. Chiplet-based design also increases the usability of chiplets in different SoCs and provides flexibility for the vendors to manufacture chiplets using any desired process technology. In this paper, we use modular SoC and chiplet-based system interchangeably. 

One of the major concerns in any network-based system is deadlock due to cyclic hold-and-wait among virtual channels (VCs)~\cite{dally1988deadlock}. Since chiplets are designed independently, their integration on an interposer brings new challenges to provide correctness validation. Connecting several deadlock-free NoCs together in the modular SoC may introduce a new kind of deadlock formed among different chiplets, as they are oblivious to each others routing algorithm~\cite{jiemingISCA2018}. 

There have been many studies that address the deadlock issue in conventional interconnection networks~\cite{dally1988deadlock,duato1993new,aniruddhISCA2018,wentzlaff2007chip,garcia2012fly,aisopos2011ariadne, flich2003applying}. Conventional deadlock avoidance techniques cannot be applied directly to modular SoCs, as they consider the whole SoC as a single network, which violates the fundamental modularity principle of the chiplet-based system design. 
Keeping modularity of the design in mind, recently, Yin et al.~\cite{jiemingISCA2018} propose Modular Turn Restriction (MTR) to impose extra turn restrictions on the boundary routers 
of chiplets to avoid  deadlocks in modular SoCs. This approach can lead to load imbalance and create several hotspots, which are detrimental for network throughput. Moreover, restrictions in MTR reduces the flexibility of SoC design.  

We exploit two key insights regarding the deadlock in modular SoC. First, since it is deadlock free inside a chiplet, deadlock in modular SoC implies that packets with destinations in the current chiplet are blocked by packets with destinations outside the chiplet. The other key observation is that packets, which are involved in deadlock, cross the boundary of the chiplets through a set of specific boundary routers. Since the chiplets and interposer have independent deadlock free routing techniques, deadlock is not possible either only inside the chiplet, or only in interposer; it must involve both. Hence, all the deadlocks formed in the system involve chiplet boundary routers.  

Based on these two observations, we propose \textbf{Remote Control (RC)}, which remotely controls injection of packets with destinations outside the chiplet to ensure deadlock freedom in modular SoCs. \textit {RC} facilitates the segregation of packets that have destinations in the current chiplet from the packets bound to outside the chiplet through a small rc\_buffer at boundary routers. The rc\_buffer, is added only in the boundary chiplet-router and it is big enough to store a whole packet bound to outside chiplet. Using \textit{RC} scheme, all the chiplet nodes ensure space availability in the rc\_buffer before injecting packets that have destinations outside chiplet. This is how we make sure that these packets will not block the VCs of the source chiplet for ever, even though they can not go outside chiplet. Interestingly, if we only handle this case that will automatically make sure that if any packet reaches its destination chiplet, deadlock free chiplet routing will ensure that they get consumed, and hence system wide deadlock-freedom is also guaranteed. 

The main contributions of this work are as follows.
\begin{itemize}
    \item We tackle an emerging problem of deadlock in chiplet-based systems using \textit{Remote Control}. We aim to provide better design flexibility to both the chiplet and SoC designers so that chiplet designers can exercise their domain specific expertise to fully optimize the chiplet performance. To the best of our knowledge, this is the first work that uses injection control based technique to guarantee deadlock freedom in modular SoC design domain.
    
    \item We formally prove and guarantee deadlock freedom in modular SoC using \textit{Remote Control}.
    
    \item We analyze existing solutions to the deadlock problems comprehensively in the context of chiplet-based systems and provide consolidated comparison. Considering several aspects together, we show that our solution is the most efficient in this context.

    \item Along with theoretical proof, we also validate the design in terms of energy, area, and timing constraints through RTL. We evaluate the performance of SoC network using network simulator and also provide system performance results using full system simulations for both homogeneous (only CPUs) and heterogeneous (CPU and GPU) system configurations. We achieve better performances than the state-of-the-art technique.  
\end{itemize}

The rest of the paper is organized as follows.
In Section~\ref{background}, we briefly describe the background, followed by a discussion on a few existing ideas that can be applied for avoiding deadlock in SoC, and motivations for \textit{RC}. Then we introduce the theory and its formal proof in Section~\ref{parking_theory}, followed by the discussion on implementation challenges. In Section~\ref{implementation}, the \textit{RC} implementation is presented. We discuss the experimentation methodology in Section~\ref{experimentation} and show the effectiveness of \textit{RC} as compared to the state-of-the-art in Section~\ref{results}. We also discuss a broad umbrella of works on network deadlock that are applied in different types of networks in Section~\ref{related}. Finally, we state our conclusions and discuss the future aspects of this work.
\section{Background}
\label{background}
In this section, we first introduce the modular SoC design concept and 2.5D active-interposer technology, as an important and elegant way of system scalability for performance boost. However, integrating independently designed chiplets introduces network deadlock, involving multiple chiplets. We study MTR and two possible conventional ways that can be applied for tackling that issue, followed by a discussion on the limitations of existing mechanisms as motivation for our work.   

\subsection{Modular 2.5D SoC Integration}
Modularity has been advocated as a new design principle to reduce the complexity and cost of SoC design. An SoC is called modular if all the chiplets on that SoC are designed independently. Contemporary multi-chiplet SoC integration use a passive silicon interposer~\cite{black2013stacking}, where the only way to make connections between chiplets is to make fixed wire connections. 

In passive interposer, dedicated wire connections are required from one chiplet to connect with different chiplets~\cite{stow2017cost}. This may lead to long wire usage with multiple repeaters, and also huge number of dedicated communication channels, making it hard to scale in terms of area and energy. In addition, the channels in the passive interposer should be standardized to make modular SoC design. Hence, there is increase in research of active interposer~\cite{jerger2014noc, kannan2015enabling, demir2014galaxy, beyne2013high, lau2014overview, vivet20153d, kim2011interposer, henry2009development} both in industry and academia. We also consider an active interposer substrate for designing the interposer network.

Active interposer facilitates interconnection between the chiplets~\cite{stow2017cost,arunkumar2017mcm,3dicorg} by adopting router design in the silicon substrate, which is more area-and energy-efficient. The integration process is generally known as 2.5D integration, featuring a silicon interposer. It is placed between the System-in-Package (SiP) substrate and the dice, where this silicon interposer has Through-Silicon-Vias (TSVs) connecting the metalization layers on its upper and lower surfaces~\cite{3dicorg}.  

\subsection{Deadlock Freedom in Modular SoC}
\subsubsection{Modular Turn Restriction (MTR)}
Based on the principle of turn restrictions~\cite{isca1992_glass_turnmodel}, recently, Yin et al.~\cite{jiemingISCA2018} propose a deadlock-free routing algorithm for modular SoC. As the best of our knowledge, this is the only work on modular SoC deadlock freedom so far. At design time, MTR finds the optimal placement and turn restrictions for boundary routers of each chiplet independently with the help of Channel Dependency Graph (CDG) analysis. The turn restrictions are applicable to both the packets that go out from the chiplet (\textit{outbound packet}) as well as to the packets that reach from other chiplets (\textit{inbound packet}). Once the list of turn restrictions is obtained, MTR applies the turn restrictions in the chiplet routing, which are applicable only for the outbound packets. For imposing turn restriction on the inbound packets, interposer routing also needs to be modified, which imposes constraint on the SoC designers and increases design complexity.

\subsubsection{VC Separation (VC-SEP)}
The idea of VC separation is applied widely to avoid protocol deadlock as well as routing deadlock based on Duato's theory~\cite{duato2001general}. We showcase it as a potential solution for SoC deadlock since it is a natural fit for this particular problem. The traffics in the Modular SoC can be categorized into two. (1) \textit{Inter-chiplet}: traffic that consists of packets, which do not have destinations in the source chiplet. (2) \textit{Intra-chiplet}: traffic that consists of packets having both sources and destinations in the same chiplet. Without any constraint on the area, cost and energy, VC-SEP naturally segregates two different traffics by providing two different virtual networks throughout the system. For outbound packets, we allocate first half of the total set of VCs, and other set of VCs are being allocated for inbound packets and all the intra-chiplet packets.

\subsubsection{In-Transit Buffer (ITB)}
The idea of ITB is originally used by Flich et al. for avoiding deadlock in irregular network~\cite{Flich:2000:PEN:335231.335235} and later extended for off-chip network in the cluster of workstations~\cite{flich2003applying}. Here, we borrow the idea of ITB and apply in Modular SoC for avoiding deadlock. ITB uses the Network Interface Card (NIC) memory as an in-transit buffer in some pre-decided nodes. Those special nodes are being selected after CDG analysis as deadlock breaking points. Any packet that reaches to those nodes are forced to eject in that node. Using DMA, the whole packet is stored in the NIC memory. In case the NIC memory gets exhausted, the packet is dropped and a NACK packet is being generated and sent to the source node for re-transmission. This process continues till the packet is ejected successfully in that special node. If the packet is successfully stored in NIC memory, that NIC sends an ACK message to the source node. In Modular SoC, to port this idea, we consider the boundary routers as special nodes and use the Network Interface (NI) connected to boundary routers to place ITB, a small buffer to store packets (no DMA) in the similar way described above. The ejection and reinjection in some special nodes break the circular channel dependency chain and hence avoids deadlock.

\subsection{Motivations for Remote Control}
Limitations in the existing techniques discussed as below motivates us to find a better solution. MTR is the state-of-the-art that identifies an important emerging problem and provides a solution. However, MTR has a few limitations. The major constrain in MTR is that extra turn restrictions can lead to non-minimal path for intra- and inter- chiplet traffics. In addition, the turn restrictions obtained by the MTR algorithm does not balance the turn restrictions among the boundaries well. Because of that, a few boundary routers get huge inter-chiplet traffic load and others do not, leading to several hotspots in the system. MTR also constrains routing design and incurs design overhead. The complexity of the CDG analysis, which is the core of this technique, gets exponentially complex with the increase in the number of chiplet routers and boundary routers, which hampers the design cycle. Moreover, MTR imposes restrictions on interposer routing to restrict the inbound packets route that increases its design complexity further.

\begin{table}[t]
    \centering
    \resizebox{\columnwidth}{!}{
    \begin{tabular}{c|c|c|c|c}
        \hline
           & Modularity & Design Efficiency & Energy Efficiency & Performance \\ \hline\hline
         MTR & ${+++}$ & $++$ & $+++$ & $--$ \\ \hline
         VCSP & $--$ & $++$ & $---$ & $---$ \\ \hline
         ITB & $+++$ & $--$ & $--$ & $--$ \\ \hline
         RC & $+++$ & $+++$ & $+++$ & $+++$ \\ \hline    
    \end{tabular}
    }
    \caption{Qualitative comparison with different deadlock avoidance techniques for modular SoC; $(+)$ means high, $(-)$ means low. We project the degree of high and low efficiency with number of $(+)$, or $(-)$, respectively.}
    \label{tab:technique_comparison}
\end{table}

Virtual Channel (VC) separation is a well known technique for avoiding deadlock. The main drawback of this approach is that it is very expensive in terms of energy and area consumption. In addition, the buffer utilization is low, which leads to sub-optimal performance. Hence, even though this solution is fairly simple, it is not attractive for designing cost-effective and high throughput modular SoC. 

\begin{figure*}[t]
\centering
    \begin{subfigure}{0.33\textwidth}
    \frame{\includegraphics[scale=0.2]{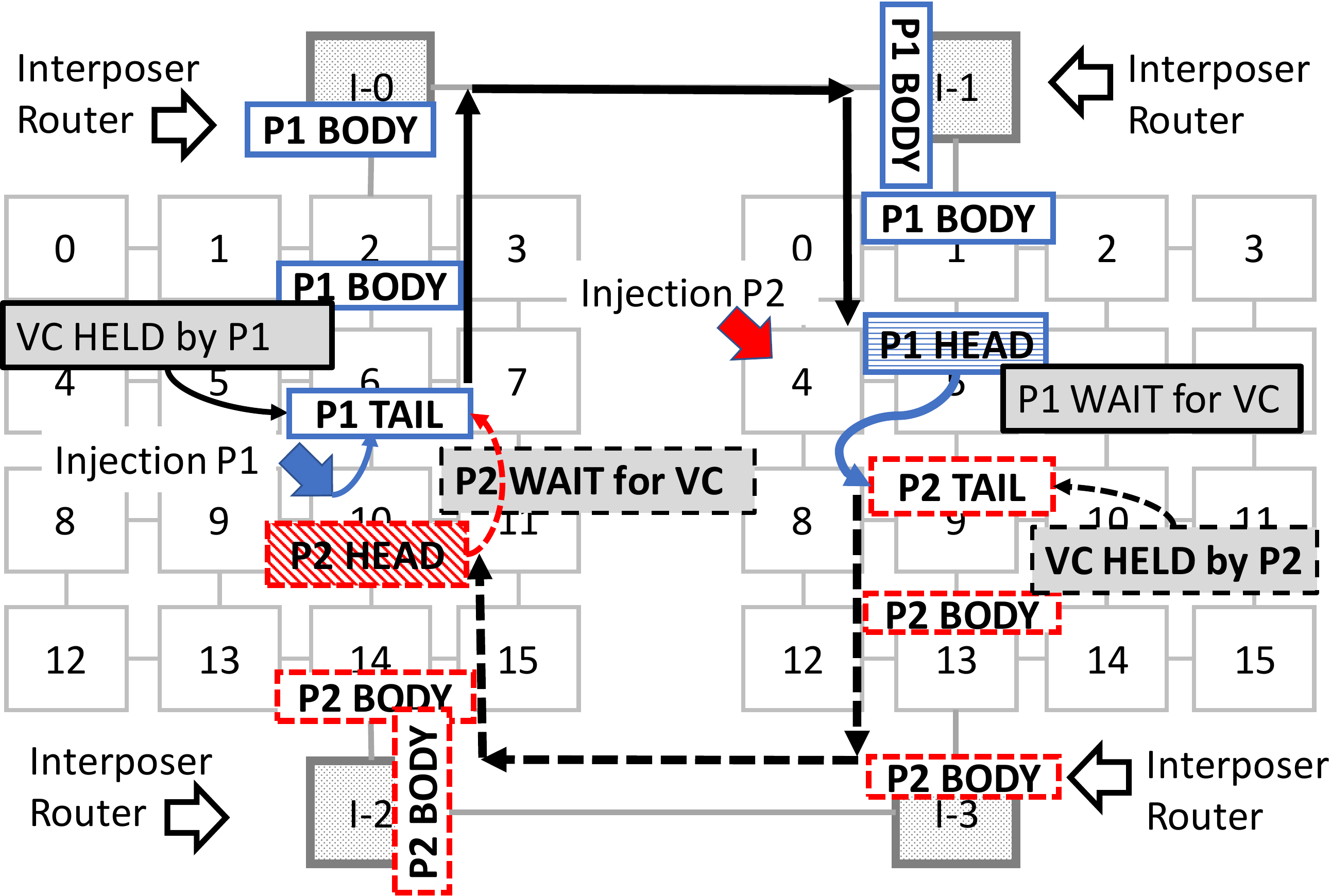}}
    \caption{}
    \label{fig:deadlock_formed}
    \end{subfigure}
    \begin{subfigure}{0.33\textwidth}
    \frame{\includegraphics[scale=0.2]{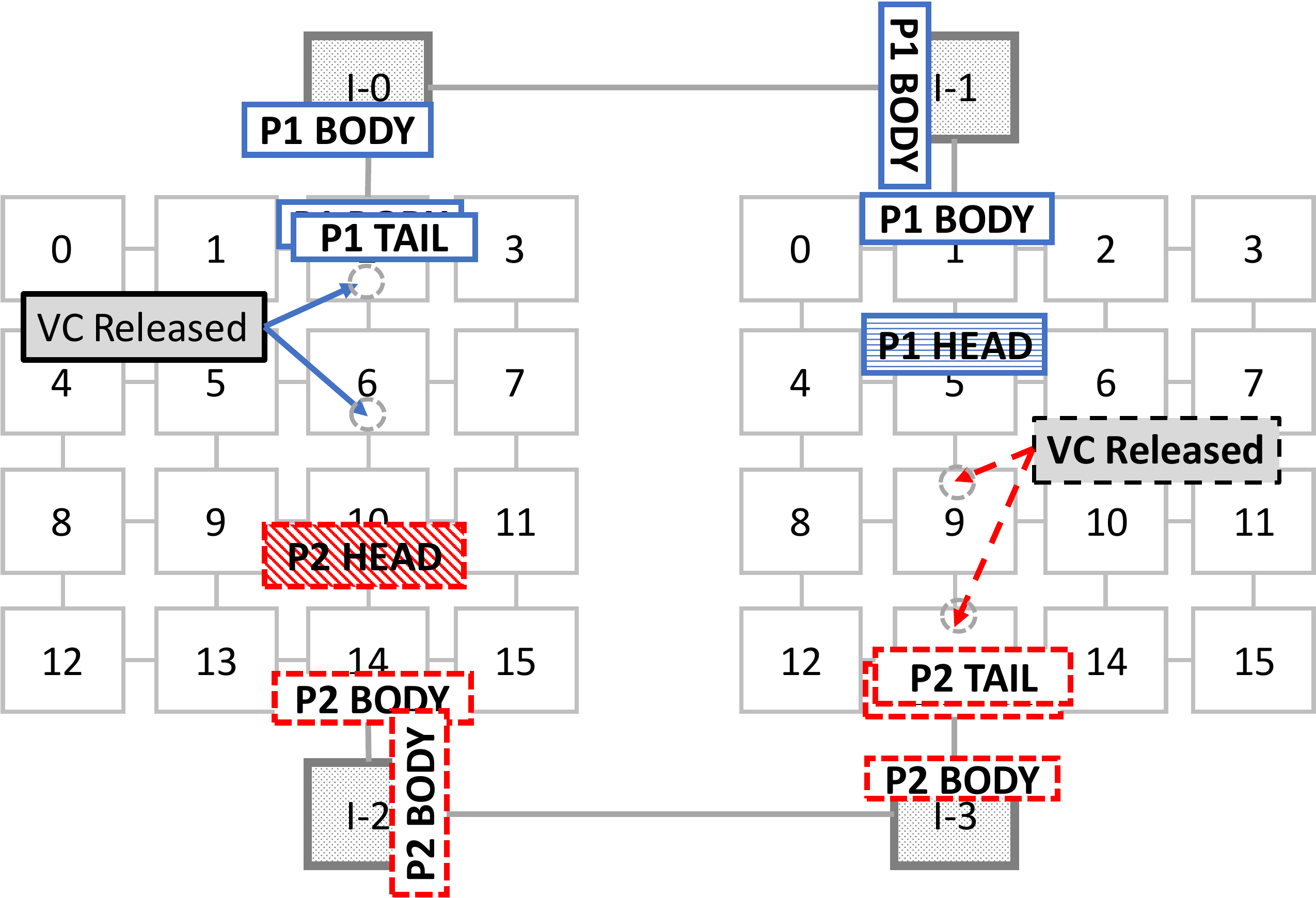}}
    \caption{}
    \label{fig:parking_introduced}
    \end{subfigure}
    \begin{subfigure}{0.33\textwidth}
    \frame{\includegraphics[scale=0.2]{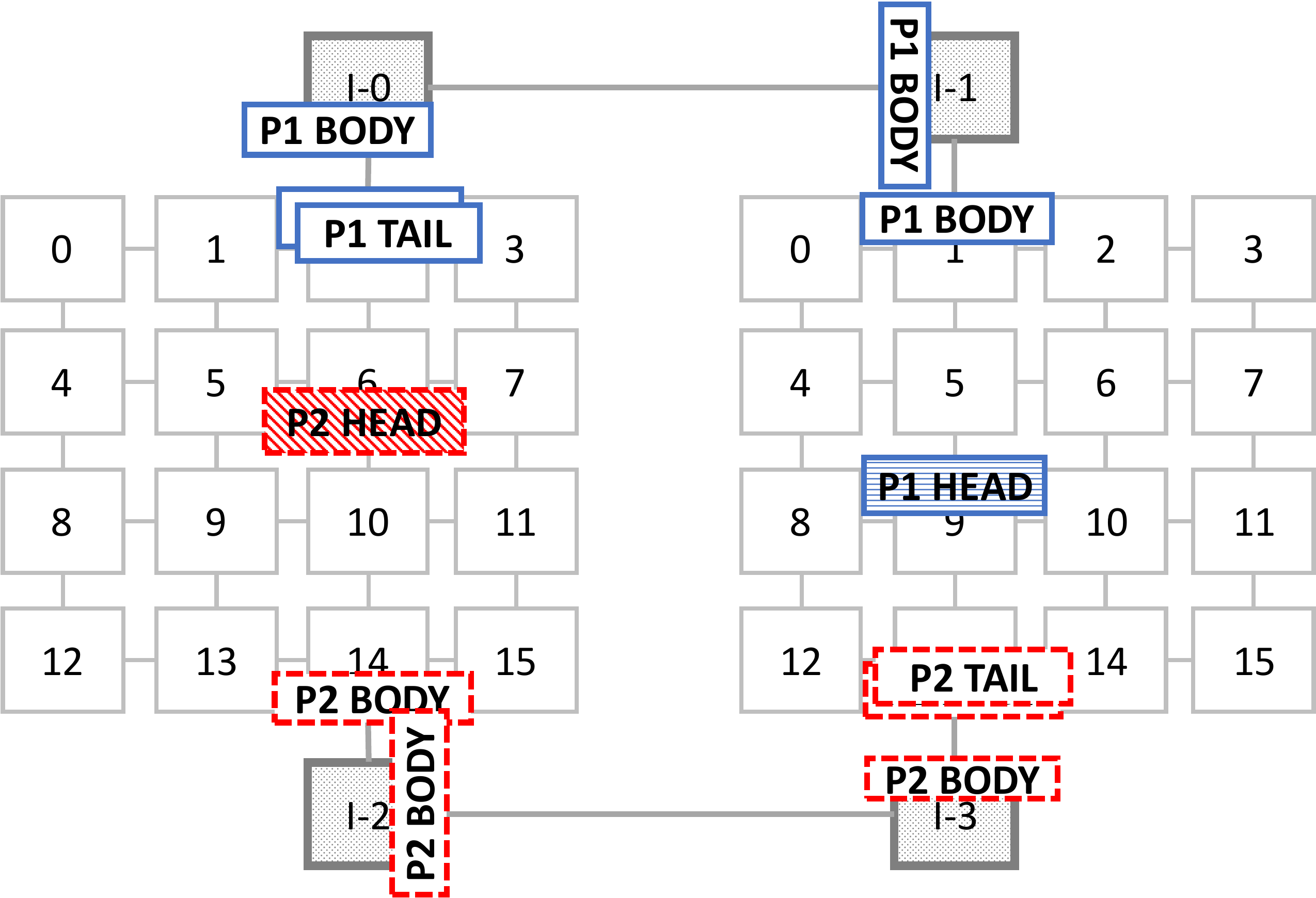}}
    \caption{}
    \label{fig:deadlock_avoided}
    \end{subfigure}
    \vspace{-0.2cm}
    \caption{\footnotesize{An example of deadlock, formed in between two chiplets (4x4 mesh), and how \textit{RC} can avoid the deadlock. (a) Packet P1 (blue solid line), and P2 (red dashed line) forming deadlock. (b) Introduced rc\_buffer with the boundary router to store outbound packets. (c) \textit{RC} avoids the deadlock by allowing all the outbound packets to be stored in the boundary router, until they get credit from downstream interposer router.}}
    \vspace{-0.2cm}
    \label{fig:deadlock_parking_example}
\end{figure*}

Use of ITB could be a promising solution for solving deadlock in Modular SoC. However, it has two major drawbacks that reduces its appeal in this context. (1) Dropping a packet in on-chip reliable network introduces unnecessary complexity and overhead. (2) Ejection and reinjection in multiple nodes increases the overall hop count, and so as the average packet latency. Furthermore, due to packet dropping/reinjection, and use of ACK and NACK packets, the overall throughput of the system also suffers. 

In Table~\ref{tab:technique_comparison}, we summarize the comparisons of
these techniques and project the expectation of \textit{Remote Control}, which aims for improving the limiting aspects of existing solutions. In a nutshell, the goal of \textit{RC} is to provide routing design flexibility and eliminate unnecessary packet dropping with packet re-transmission by introducing a flow control based technique. Additionally, \textit{RC} targets to save energy and area by segregating traffics only in chiplet boundary routers.
\section{Remote Control}
\label{parking_theory}

In this section, we first walk through a simple practical example to show a case of deadlock formation between two chiplets and how \textit{RC} can solve it. Then we generalize it for any chiplet-based systems and theoretically prove that \textit{RC} guarantees deadlock freedom in Modular SoCs.

\subsection{Deadlock in Modular SoC}
\label{sec:deadlock}
Figure~\ref{fig:deadlock_formed} shows a deadlock case in a modular SoC, where two 4$\times$4 2D mesh chiplets are connected through an interposer. We denote router $i$ on chiplet $j$ as R-$i$/C-$j$ for simplicity, where chiplet-0 is on the left and chiplet-1 is on the right. In this system, R-2/C-0, R-14/C-0, R-1/C-1 and R-13/C-1 are boundary routers connected to interposer network.  
Packets P1 and P2 are two outbound packets (packets that have source and destination in C-1, and C-0, respectively) are in a circular hold-and-wait situation, forming a deadlock. The P2-head flit in R-10/C-0 requests for the south VC of R-6/C-0, which is held by packet P1. On the other hand, P1-head flit in R-5/C-1 requests for the north VC of R-9/C-1, which is taken by P2. Such a case creates a circular hold-and-wait situation and forms a deadlock, where neither P1, nor P2 can make forward progress\footnote{Progress/forward-progress means moving near to the destination.}. To avoid deadlock in this scenario MTR may impose turn restriction from {\it R-6/C-0 : R-2/C-0 : I-0}, increasing pressure on the other boundary of C-0 for outbound traffic. Note that MTR needs to impose more turn restrictions to avoid all other possible circular hold-and-wait scenarios. 

\subsection{Deadlock Avoidance using \textbf{\textit{\large RC}}}
As shown in Figure~\ref{fig:deadlock_formed}, outbound packet P1 in C-0 is blocking P2 packet to reach its destination. P2 is blocking P1 from getting consumed in C-1. \textit{RC} separates outbound packets from others in the boundary router, and allows them to be stored completely in the rc\_buffer until they get a credit back from the downstream interposer router. This makes sure that the chiplet VCs will get free in bounded time, once the header flit reaches the rc\_buffer. Hence, P1 will release all the VCs that are currently blocked in C-0, and so do P2 release VCs in C-1. That is why the circular channel dependency among chiplets will never result into deadlock. 
In the following section, we generalize the whole system into two nodes to prove the deadlock freedom guaranteed by \textit{RC} for any modular SoC.  

\subsection{Proof of Deadlock Freedom}
\begin{figure}[h]
    \centering
    \includegraphics[scale=0.19]{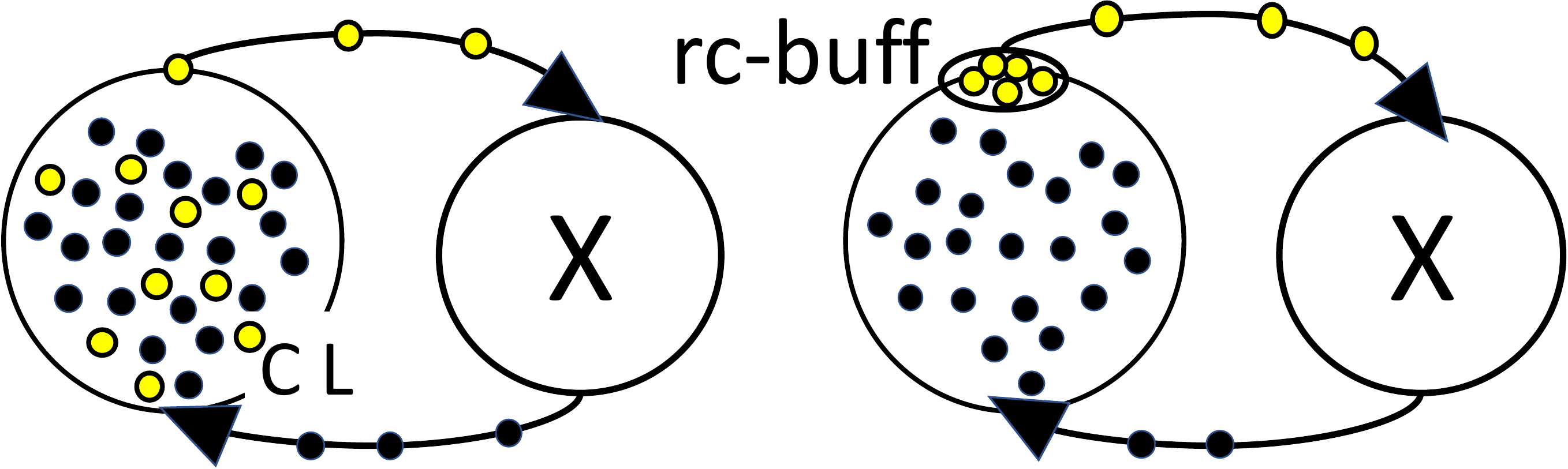}
    \caption{Hold-and-Wait dependency chain.}
    \label{fig:hold_and_wait}
\end{figure}

To simplify the problem, we abstract a given chiplet as \textit{CL}, the rc\_buffer attached to \textit{CL} as \textit{rc-buff}, and the rest of the SoC network as \textit{X}, as shown in Figure~\ref{fig:hold_and_wait}. The solid black dots in CL represents the packets that have destination in the current chiplet, and yellow dots are outbound packets that have destination in \textit{X}.

\begin{figure*}[t]
    \centering
    \includegraphics[scale=0.54]{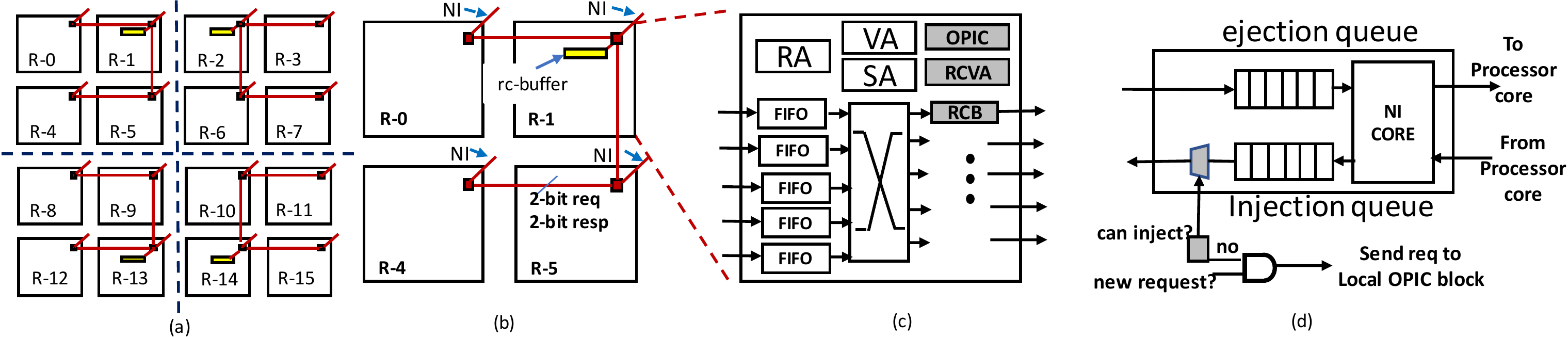}
    \caption{Remote Control: (a) Outbound Injection Control System (OPIC) consists multiple OPIC blocks connected in a tree fashion. Each router has one OPIC block. (b) One OPIC tree, and each edge is 2-bit request and response line. (c) Boundary router with the newly added components marked with gray. Note that the router attached to a non-boundary node does not have RCVA and RCB. (d) The changes in NI of the non-boundary router marked in gray. There is no change in NI attached to boundary routers.}
    \label{fig:overall_rc}
\end{figure*}

 There are two assumptions for \textit{RC} as follows:
\begin{itemize}
    \item \textbf{Assumption 1.} All the outbound packets in \textit{CL} need to go through \textit{rc-buff} to reach \textit{X.}
    \item \textbf{Assumption 2.} \textit{rc-buff} has sufficient space to store all the in-flight outbound packets of \textit{CL}.
\end{itemize}

\textit{We prove deadlock freedom of \textit{Remote Control} by contradiction. As shown in Figure~\ref{fig:hold_and_wait}, \textit{CL} can be involved in deadlock with circular hold-and-wait dependency chain (left figure). Let us assume \textit{CL} can also be involved in a deadlock even after introduction of \textit{rc-buff} in right figure. According to the necessary and sufficient Coffman conditions for deadlock~\cite{csur1971_coffman_deadlock}, there must be hold-and-wait for \textit{CL} $\rightarrow$ \textit{rc-buff}, meaning \textit{rc-buff} does not have enough space for the in-flight outbound packets in \textit{CL}, which contradicts \textit{Assumption 2}. Therefore, by avoiding circular hold-and-wait dependency chain as shown in Figure~\ref{fig:hold_and_wait} (b), we prove that \textit{RC} guarantees deadlock freedom in modular SoC.}
\subsection{Implementation Challenges}
\label{challenges}
We build the theoretical proof based on two assumptions. Major implementation challenge is to realize them in practical implementation. We assume that all the outbound packets in \textit{CL} need to go through rc\_buffer to reach to \textit{X}. We know that all the outbound packets go through boundary routers. Hence, we place the rc\_buffer in the boundary router. The challenge would be to place and implement rc\_buffer such that, it does not introduce extra delay.

According to the second assumption we need to make a way that the boundary router can store all the outbound packets completely if they can not go to the downstream interposer router. Since, it is practically impossible to provide space for any number of outbound packets, we need to control their injection at the non-boundary routers, depending on the space availability in rc\_buffer. The challenge would be to get the information about space availability in rc\_buffer, situated in boundary router, while a non-boundary router tries to inject an outbound packet. This information should be propagated with minimum delay as the outbound packet might increase the queuing delay for the following packets stored in the same injection queue.

\subsection{Routing Oblivious Design}
We propose injection control based technique that is independent of both the chiplet routing and the interposer routing. The boundary routers are considered as local destinations for the outbound packets as they need to reach there first for going out of the chiplet. Once the local destination is decided then any routing technique can be used to reach to the boundary router. Through the boundary router when the packet reaches the downstream interposer router, it follows the interposer routing to reach the local destination in the interposer. Since the interposer takes care of the communication between two chiplets, one inter-chiplet packet only traverses through its source and destination chiplets. Once the packet enters the destination chiplet, it follows the routing of that chiplet to reach to the destination node in that chiplet.
\section{Implementation}
\label{implementation}
For implementing \textit{RC}, we modify (1) chiplet routers, and (2) Network Interface (NI) of the non-boundary nodes. In the non-boundary routers, we need to implement injection control logic that regulates the injection of the outbound packets from its injection queue in NI, whereas in the boundary routers, we need to segregate the outbound packets from the other packets. In this section, we  discuss about the implementation details of \textit{RC}, followed by a short discussion on the integration aspects of \textit{RC} in modular SoC.

\begin{figure}[h]
    \centering
    \includegraphics[scale=0.4]{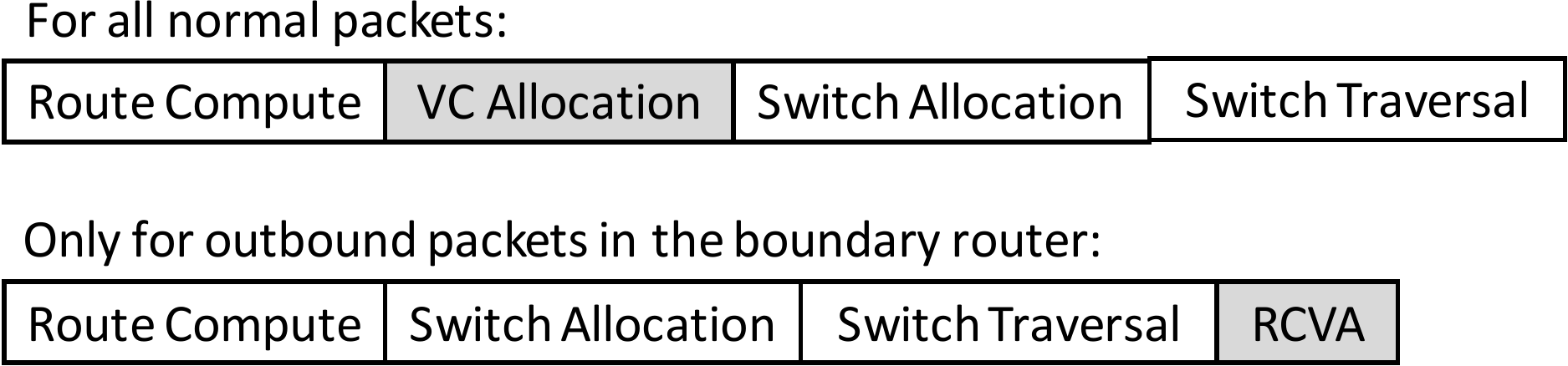}
    \caption{Router pipeline stages in case of normal packets followed by modified router pipeline in case of handling outbound packets in the boundary router.}
    \label{fig:router_pipeline}
\end{figure}

\subsection{Boundary Router}
\subsubsection{RC VC Allocation and rc\_buffer}
Each boundary router has three new components (1) \textit{RC} VC Allocation (RCVA), \textit{RC} buffer (RCB), and OPIC) as shown in Figure~\ref{fig:overall_rc}c, marked gray. RCVA implements two functions. (1) It makes sure that the outbound packets that are not being injected in that boundary router, do not participate in VC allocation (VA) and directly do switch allocation. We bypass VA stage of the router and save the latency for that stage. If the switch allocation is succeeded, then the packet reaches the output side through the crossbar. On the output side of the crossbar, we make an entry for that packet in RCB. RCB is a collection of FIFO queues. Once head-flit of a packet arrives, RCB makes an entry for the whole packet reserving one of the FIFO so that other flits follow the head flit. The depth of the FIFO must be the maximum number of flits in any packet. Availability of a FIFO queue is already assured by OPIC, discussed in details later. (2) Before making entry in RCB, in any cycle, RCVA checks if there is any candidate that is waiting in RCB for VC allocation. The VC allocation logic in RCVA is much simpler and straight forward than that in the VA, as RCVA deals with only one port. Moreover, since RCB collects all the outbound packets from all the input VCs, VA does not deal with the outport that connects the downstream interposer router. RCVA does not increase the number of stages in the router for any kind of packets. For outbound packets we consider a different router pipeline, which has same number of router stages as in the normal router as shown in Figure~\ref{fig:router_pipeline}.

\textbf{\textit{Why we interchange the positions of the router pipeline stages?}} We allow the outbound packets to reach to the rc\_buffer irrespective of the VC state in the downstream interposer router, unlike in normal router pipeline where a packet is allowed to perform switch allocation only after successful reservation for downstream VC. By appending the RCVA at the end of the pipeline, we ensure the input VCs are reserved only for a finite duration by the outbound packets.

\textbf{\textit{How much is the overhead of RCVA?}} None. First of all RCVA is only selectively used by outbound packets in the boundary routers. Outbound packets do not participate in regular VC allocation, which saves cycles. Moreover, since RCVA is very lightweight it takes much lesser time than VC Allocation. Conservatively we consider one cycle for VC Allocation and also for RCVA. In addition, since the link is dedicated only to RCVA, as soon as the credit is available, one flit is sent to the downstream interposer VC in each cycle.       

\textbf{\textit{Is it possible to use existing output buffers in the router in place of RCB?}} Yes. If in any router that uses output speedup, we can reuse the existing output buffer with a few additional control logic. For instance, we need to connect that with the local OPIC block.

\begin{figure}[t]
    \centering
    \includegraphics[scale=0.34]{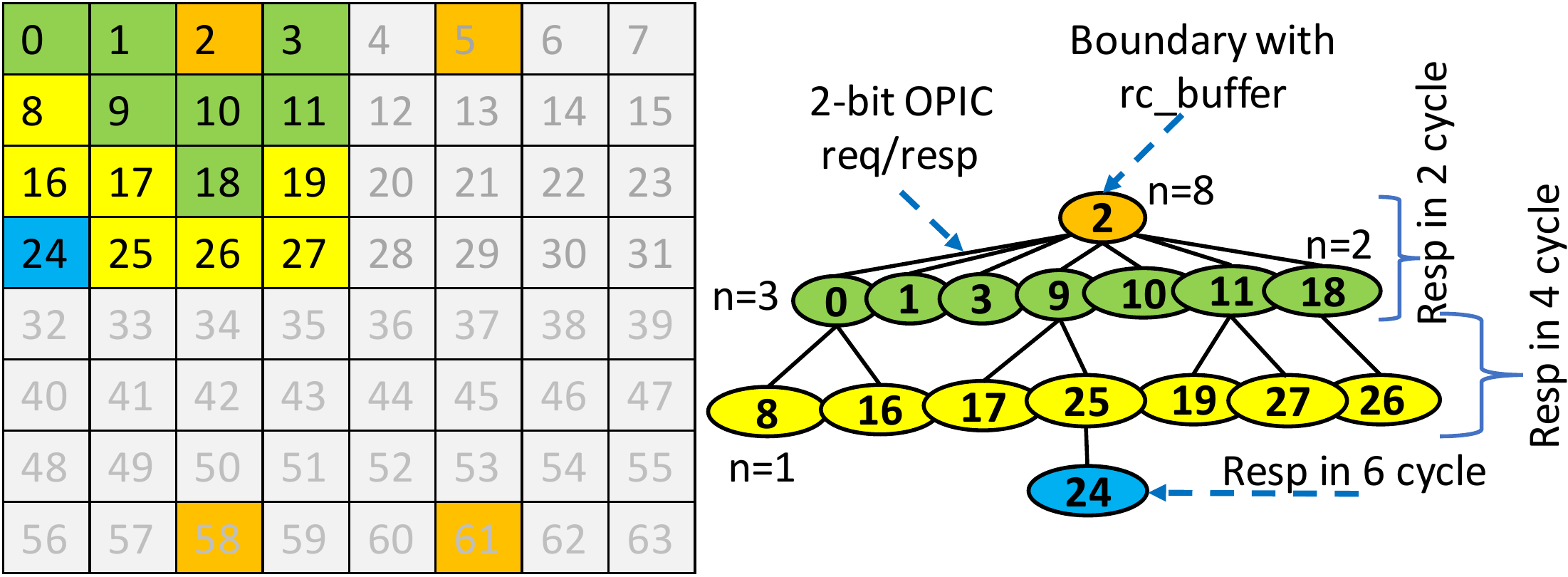}
    \caption{{\footnotesize Example of OPIC connection in ${8\times8}$ mesh with 4 boundary routers with rc\_buffer connected. The {\it n} referred here is the same in Figure~\ref{fig:opic_block}}.}
    \label{fig:opic_connect_example}
\end{figure}

\begin{figure}[t]
    \centering
    \includegraphics[scale=0.43]{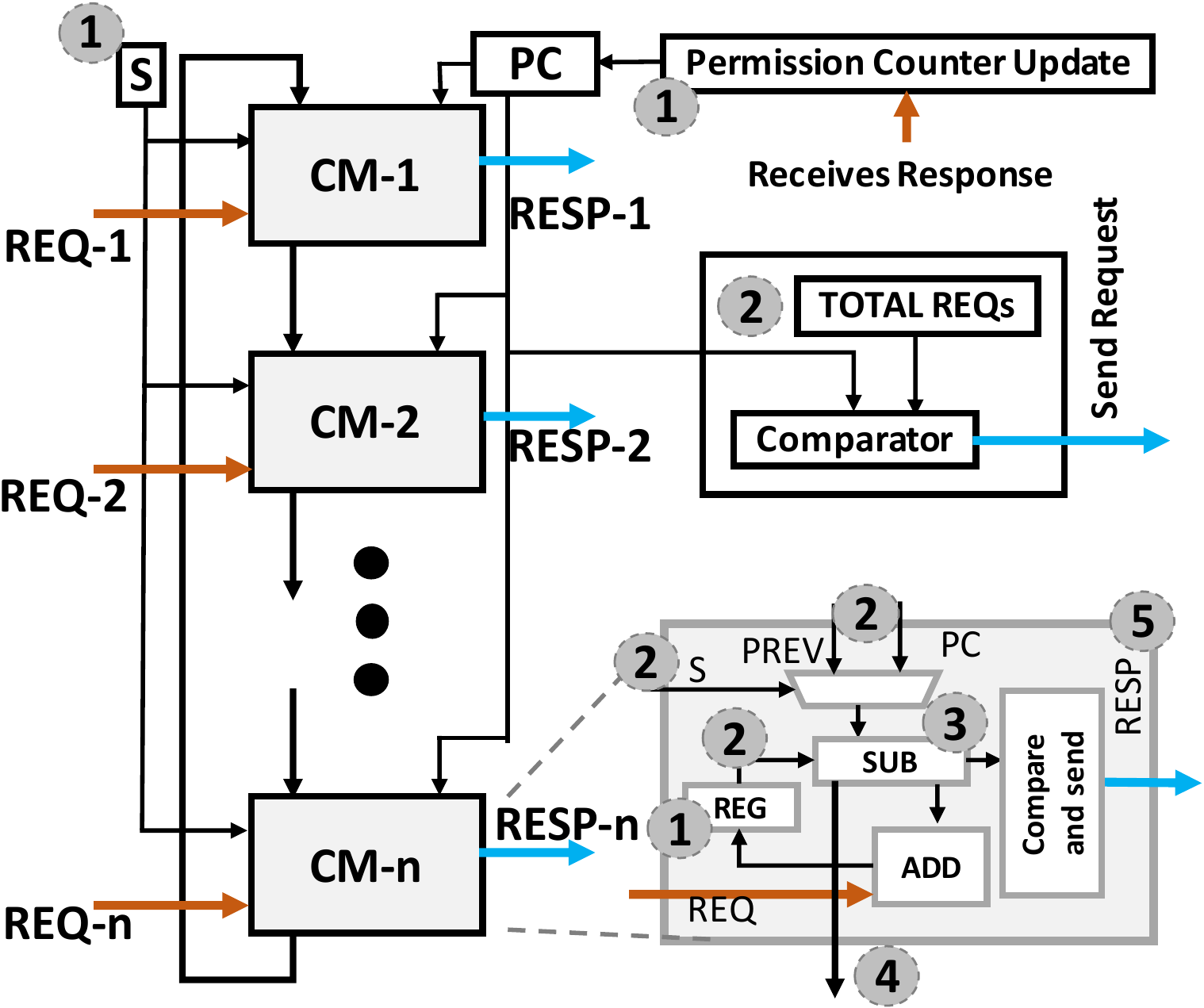}
    \caption{{OPIC block logic that consists of compute modules (CM), a Permit Counter (PC), CM selection logic (S), Request (REQ) sending block, and PC update block. CMs are connected in a ring, and CM selection logic selects a CM in Round Robin fashion in each cycle as a starting point of computation in the ring. The PPC value is read only by the starting point and rest CMs read previous CM's output (PREV). Computation in the ring is completely combinational, as we can read registers (REG) asynchronously. The ADD and SUB units do multi-bit addition and, difference calculation respectively. SUB block decides the minuend and subtrahend and sends residue to the next CM accordingly. Compare-and-Send module sends response (RESP) signal if any/all of the request(s) is/are permitted. "n" represents the sum of number of children connected and number of NIs in that node.}} 
    \label{fig:opic_block}
\end{figure}

\subsubsection{Outbound Packet Injection Control (OPIC)}
We first walk through an example to understand the OPIC system that consists of connection of OPIC blocks in a tree fashion. Then we describe the design of the OPIC block in detail, followed by a discussion on its design choices, feasibility and applicability to a bigger system.

Figure~\ref{fig:opic_connect_example} shows an example of OPIC connection in an ${8\times8}$ chiplet. Node-2 is the boundary and the OPIC block in node-2 connects with OPIC block in node-{0, 1, 3, 9, 10, 11, 18} using request and response lines of width 2-bit each in a n-ary tree. Similarly, node-0 is connected with node-{8, 16} with the request and response lines, and so on. For instance, at any cycle {\it t}, node-8 and node-27 want to inject an outbound packet, the requests will be registered in node-0 and node-11 at the beginning of {\it t+1}, respectively. At {\it t+2} the requests from node-0 and node-11 will be registered in node-2. Let us suppose the rc\_buffer does not have a space, in that case the requests will be standby in these nodes. Now suppose at cycle {\it T}, one packet space gets free in rc\_buffer and depending on the arbitration one of these two will get the response at {\it T+1}. Suppose node-0 got the response, then at cycle {\it T+2} node-8 will get the permission to inject the outbound packet. 

An OPIC block is attached to the router of each non-boundary chiplet node to control the injection of outbound packets from the NI of that node. Figure~\ref{fig:opic_block} shows the main components of the OPIC block and the logical connections among them. OPIC blocks are connected in a tree like network (root is the node with rc\_buffer) using their multi-bit REQ and RESP lines as shown in Figure~\ref{fig:opic_connect_example}. Each Compute Module (CM) in OPIC block is responsible for handling \textit{RC} REQ and RESP for each child block. The Round Robin CM-selection-logic $(S)$ is implemented using a counter and decoder. \circled{1} The counter is incremented in each cycle and reset to zero once reaches maximum value. The decoder decodes the count and sends a high signal to one of the CMs. Along with counter increment in the beginning of a cycle, PC update and REG update also take place. PC is updated using RESP received in the last cycle. In REG, we record the number of REQs received in the last cycle. We also calculate the total number of REQs using ADD block (this addition is not shown in Figure). \circled{2} If the PC value is less than the total number of REQs, it sends the difference to the parent OPIC block. The SUB blocks in CMs calculate the difference between REG, and PREV or PC depending on the CM selection ($S$) signal status. \circled{3} Compare and send block checks whether the REQ for that particular requester is served or not. \circled{4} The remaining rc\_buffer permissions, after the SUB is forwarded to the next CM in the ring. \circled{5} Finally the RESP is sent to the requester child node. 

It is evident that the complexity is maximum in the root node (there are 8 CMs in the ring) and then maximum 2 in any other node. Our hardware synthesis shows the computation delay for 8-CM-ring preserves setup-time and hold-time assuming 2GHz network clock frequency. We also add the wire delay to check whether signals are reaching on time. In case the number of CM increases for a different topology, instead of serving request and responses in one cycle, it can be processed in more than one cycle breaking the computation in a pipelined fashion. The current implementation takes 1 cycle for REQ send-and-register, and 1 cycle for process-and-send RESP at each hop. Each parent can send RESP to all its children in the same cycle and can receive REQ from all of them in the same cycle. 
Therefore, even with more levels in OPIC system tree for a bigger network, only constant latency is added in the overall OPIC delay.
Moreover, OPIC tree is independent of chiplet topology as it only establishes REQ/RESP channels between a node and its corresponding boundary router for sending outbound packets.

\subsection{Non-boundary Routers}
In a non-boundary router, we append a control on injection process, which allows all the intra-chiplet packets to go without any check. Only for inter-chiplet traffic, the modified injection system checks for injection permission from local OPIC block. In Figure~\ref{fig:overall_rc}d, we show that if the permission is not there, then the outbound packet is not injected and a request for permission is sent by the local OPIC block to the remote OPIC block in the boundary router through OPIC tree. Once the response reaches, the outbound packet is injected.

\begin{figure}
    \includegraphics[scale=0.26]{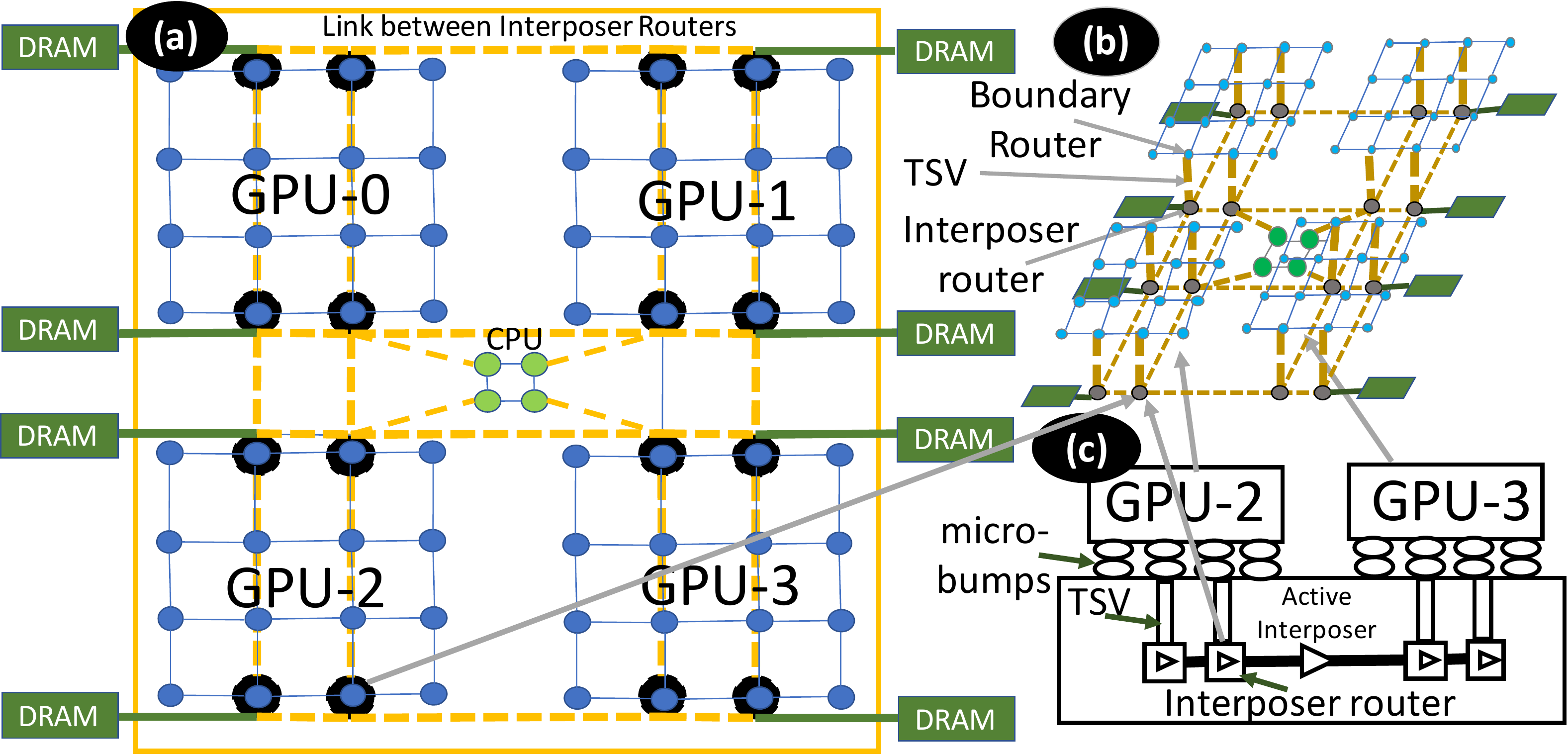}
    \caption{\footnotesize{SoC viewed from different angles. (a) Shows the top view of the SoC. There are four GPU chiplet (${4x4}$ mesh) at the four corners, and a CPU chiplet (${2x2}$ mesh) in the center. DRAM memory is connected with edge interposer routers. (b) 3D view of the same SoC, highlighting the interposer router, boundary router, and TSV that connects them. It also show the active interposer, and the mesh network in the interposer. (c) Microscopic cross-section view of SoC highlighting the micro-architectural details of 2.5D SoC integration on active interposer.}}
    \label{fig:soc_network}
\end{figure}

\begin{figure*}[t]
    \includegraphics[scale=0.31]{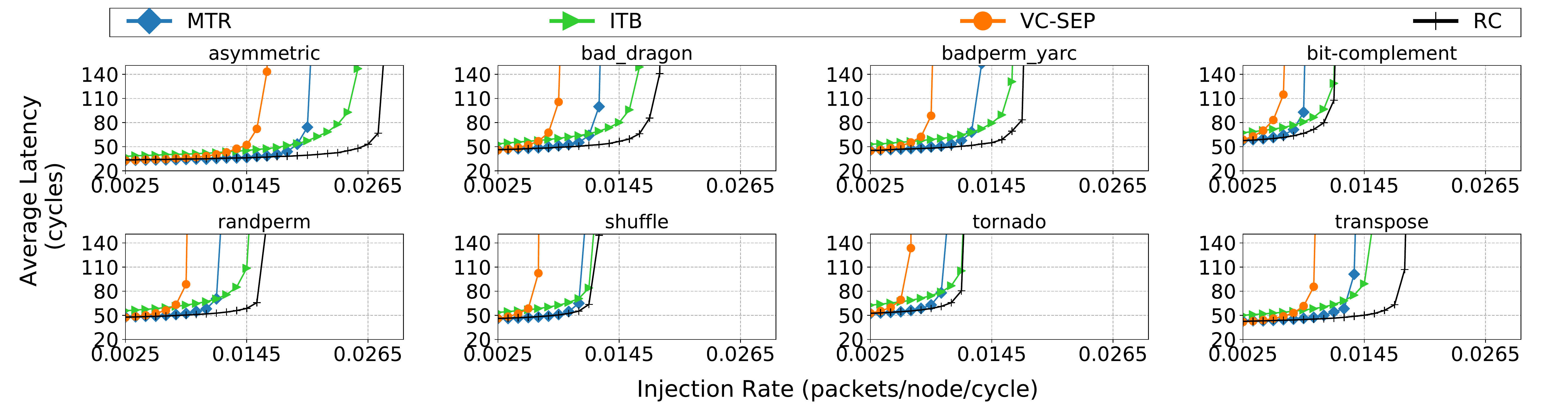}
    \vspace{-0.1in}
     \caption{\footnotesize{Throughput graph for synthetic traffic pattern study. (\#VC = 2, VC buffer size = 4, packet size = 8 flits, rc\_buff = 4 packets). }}
     \label{fig:throughput_graph}
     \vspace{-0.1in}
\end{figure*}

\begin{figure}[t]
    \begin{subfigure}{0.11\textwidth}
    \includegraphics[scale=0.21]{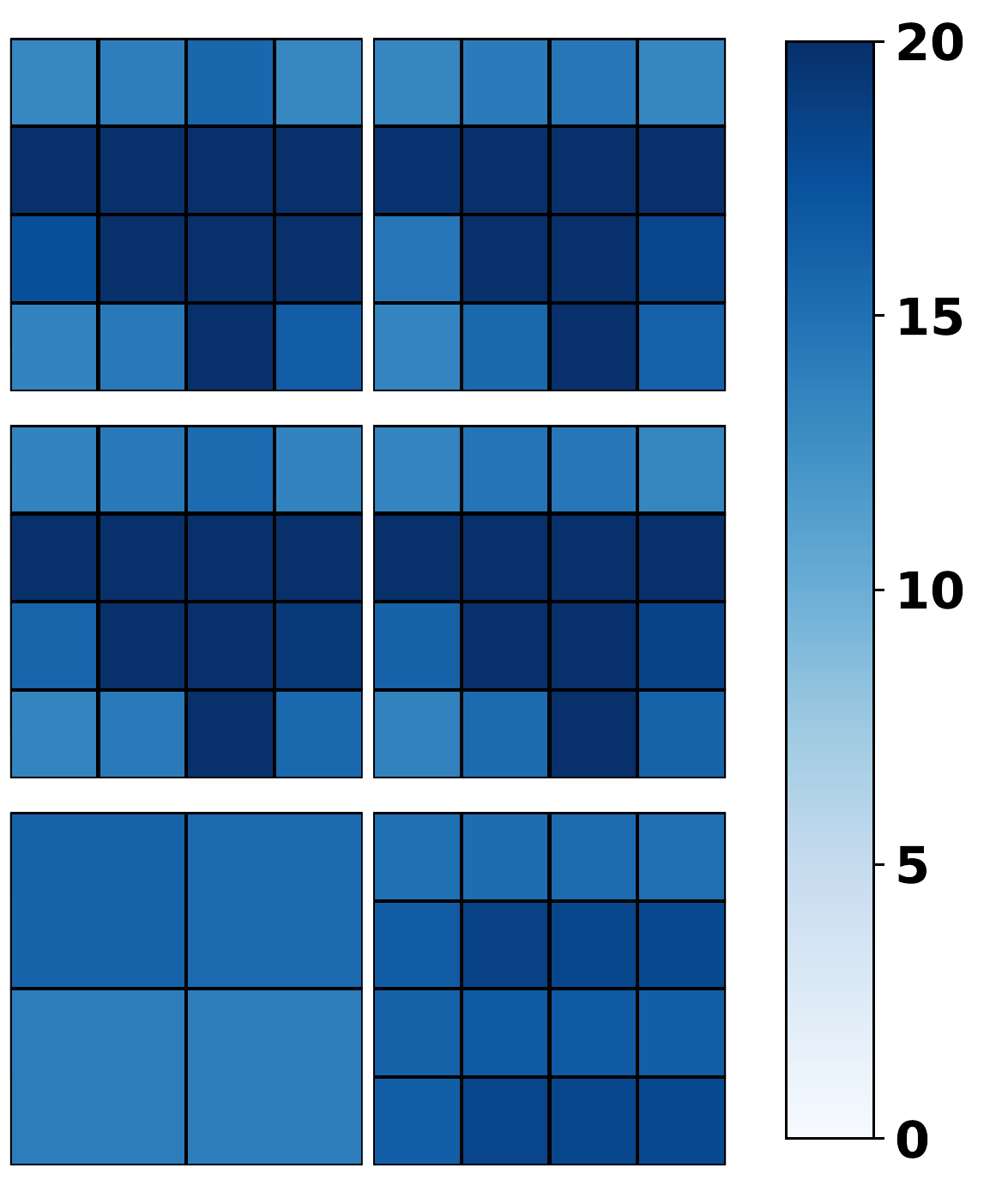}
    \caption{\footnotesize{MTR.}}
    \label{fig:modular_hotspots}
    \end{subfigure}
    \begin{subfigure}{0.11\textwidth}
    \includegraphics[scale=0.21]{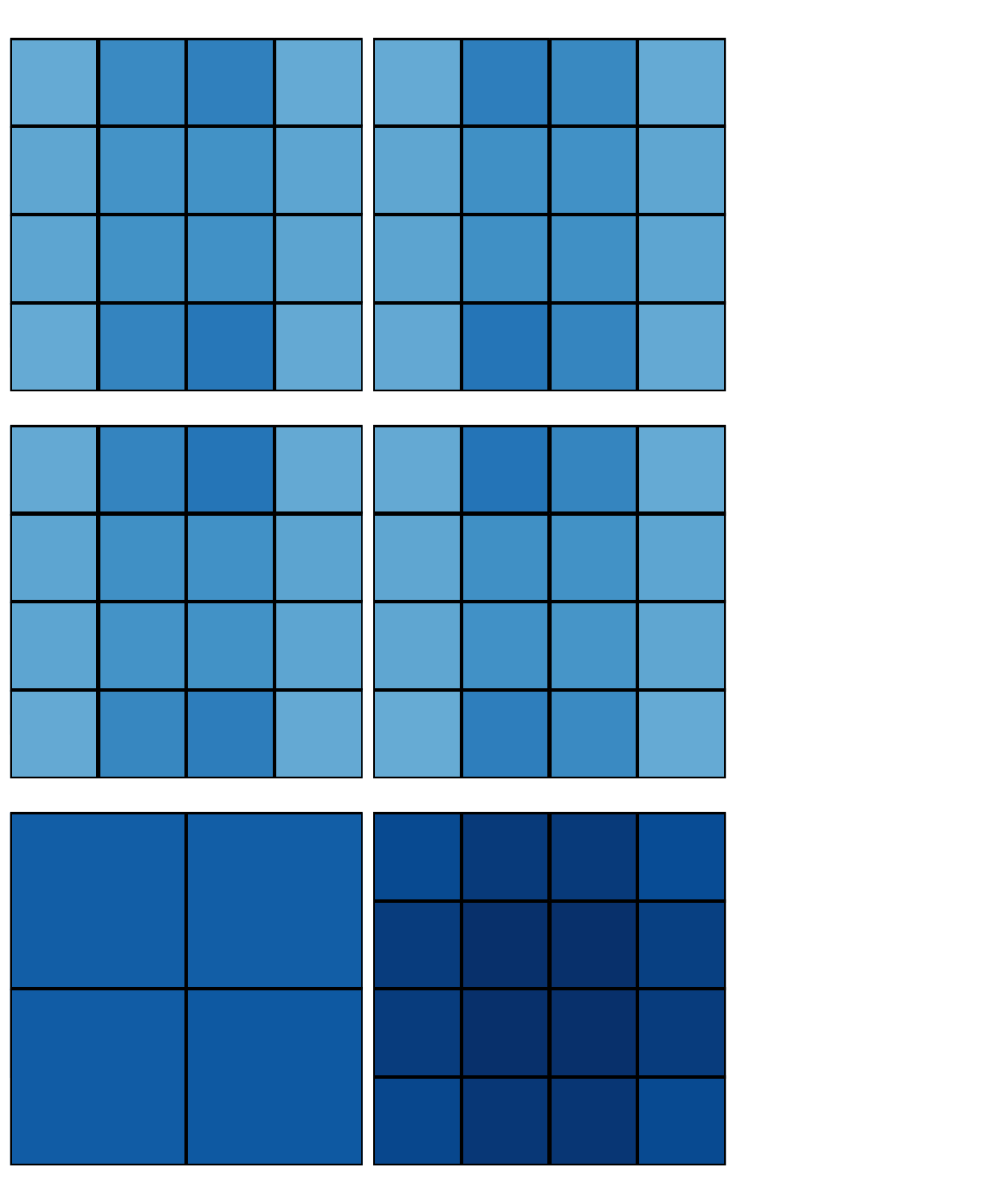}
    \caption{\footnotesize{ITB.}}
    \label{fig:itb_hotspots}
       \end{subfigure}
    \begin{subfigure}{0.11\textwidth}
        \includegraphics[scale=0.21]{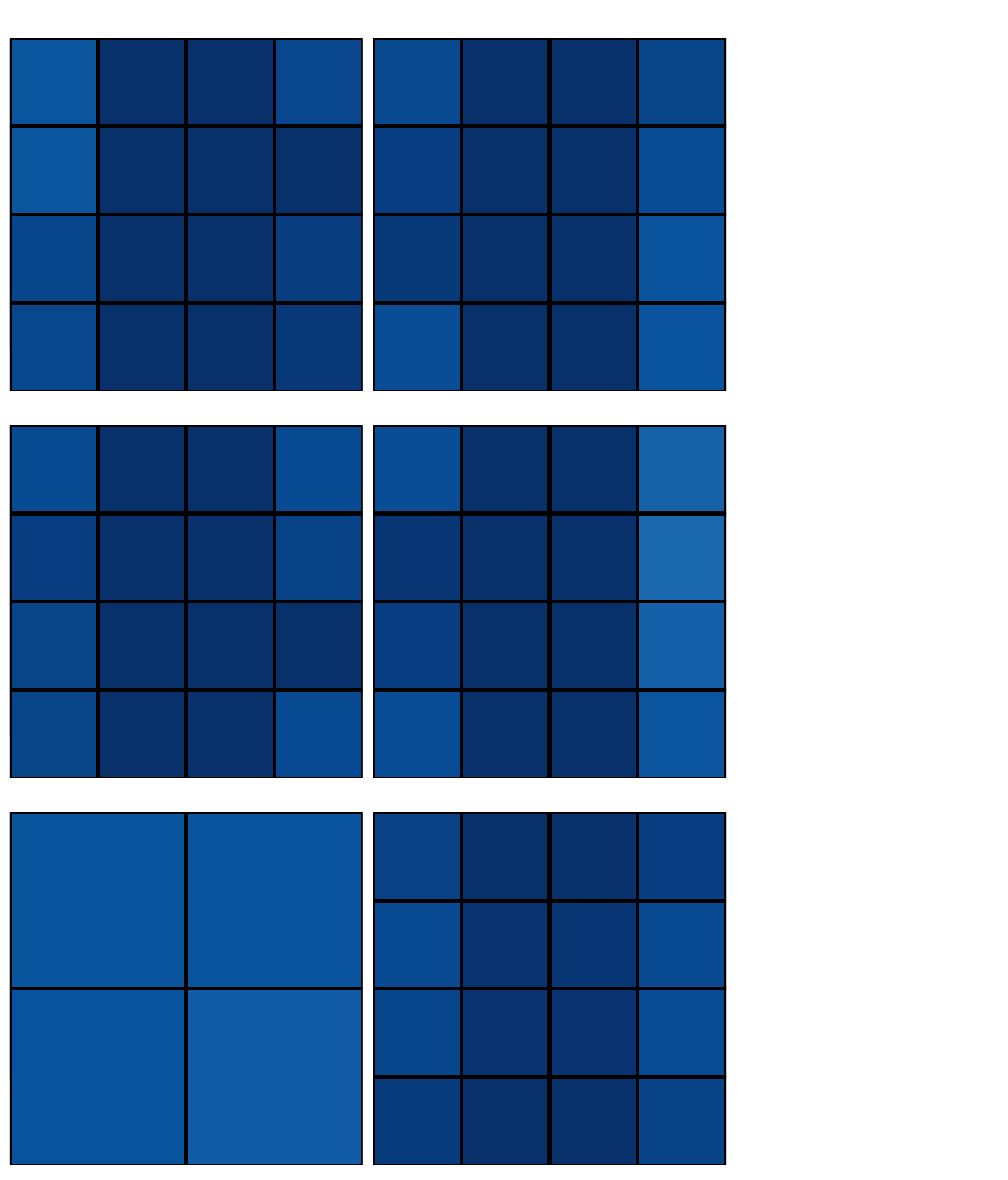}
    \caption{\footnotesize{VC-SEP.}}
    \label{fig:vc_hotspots}
    \end{subfigure}
    \begin{subfigure}{0.1\textwidth}
        \includegraphics[scale=0.21]{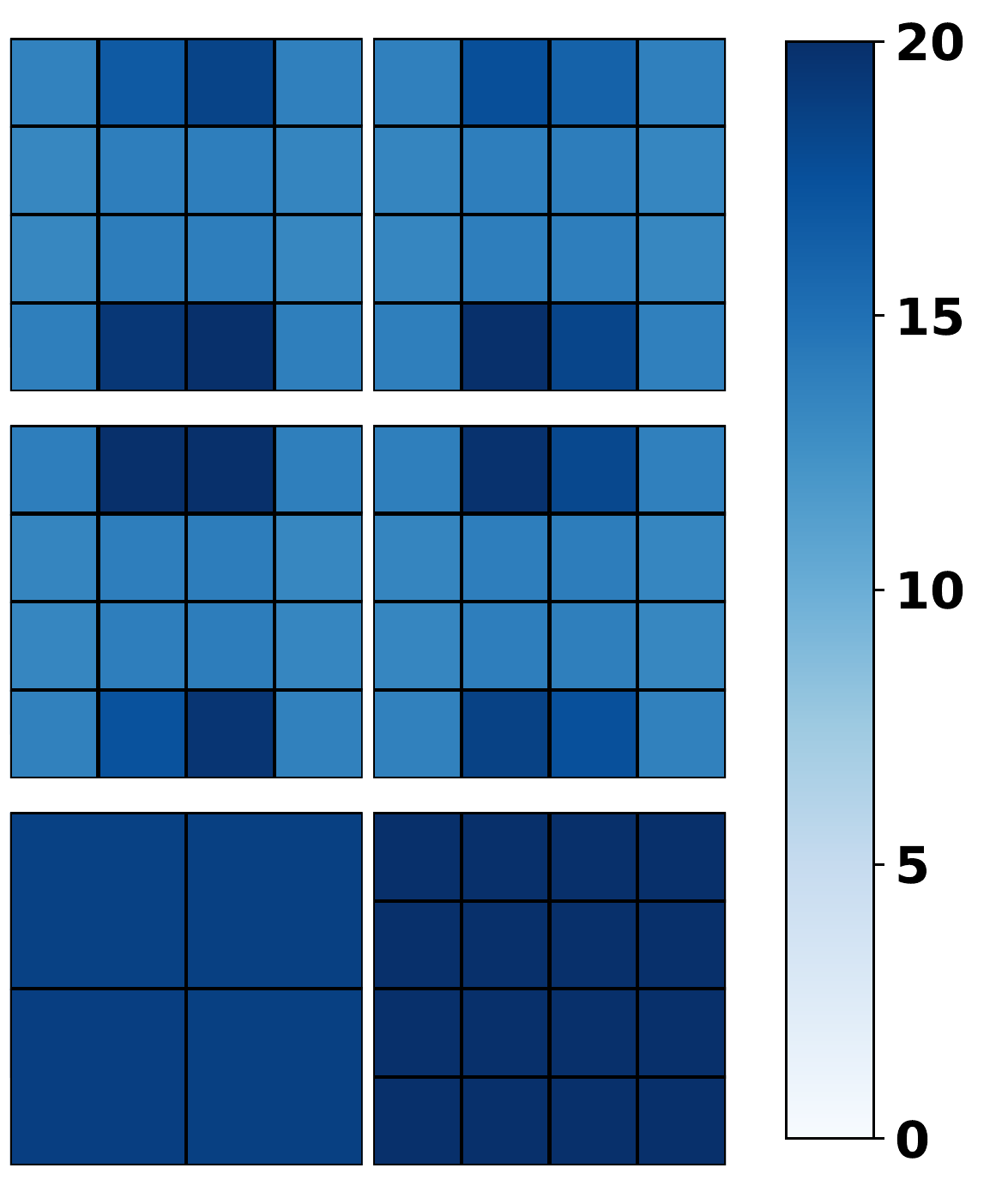}
    \caption{\footnotesize{RC.}}
    \label{fig:rc_hotspots}
    \end{subfigure} 
     
    \caption{\footnotesize{Heat map of average packet residency latency (cycles) per router (smallest cube) in UR plotted for \textbf{near saturation point} for each technique. Top four cubes (big cubes) represent GPU chiplets, having 16 routers (small cubes) each. The bottom-left cube is the CPU chiplet having 4 routers. Beside the CPU chiplet we show silicon interposer with 16 interposer routers.}}
    \label{fig:hotspots}
\end{figure}

\begin{figure}[t]
    \centering
    \includegraphics[scale=0.6]{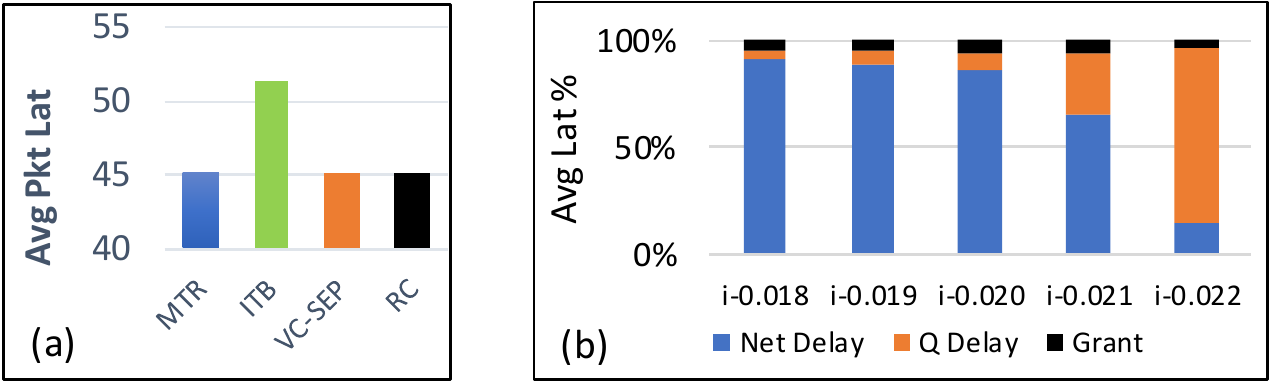}
    \caption{Analysis on throughput graph in Figure~\ref{fig:throughput_graph}. (a) Zero load latency of UR, representing the trend for others as well. (b) The network breakdown for \textit{RC} across different injection rates for UR. "Net Delay" is the network delay including retry latencies. "Q Delay" is the injection queue latency. "Grant" response latency for getting the permission from OPIC including waiting for rc\_buffer full condition.}
    \label{fig:zero_load_latency_break}
\end{figure}

\subsection{Integration in Modular SoC}
Once the chiplets are equipped independently with \textit{RC}, they are ready to be integrated in an SoC using 2.5D active interposer as shown in Figure~\ref{fig:soc_network}. Since \textit{RC} allows complete flexibility of designing an SoC, we choose one simple SoC design~\cite{jiemingISCA2018} to validate our technique. In this SoC, the interposer provides a ${4\times4}$ mesh network for inter-chiplet communication. Since in the interposer network, edge routers are connected with DRAM memory, memory controllers are connected with those routers. The edge routers in the interposer, also contain the coherence directory. 
Chiplets are connected with the interposer using micro-bumps. TSV connects the micro-bumps with the interposer routers. Interposer routers use internal link to connect with each other. For instance, in Figure~\ref{fig:soc_network}c, if GPU-2 wants to send a request packet to GPU-3, then that packet will reach to the boundary router of GPU-2 first. In the boundary router, the packet will make an entry in rc\_buffer. From the boundary router, the packet will reach to the interposer router through the TSV. Once the packet reaches to the interposer router, it will be routed to the interposer router that is connected to GPU-3. Again, through the TSV the packet will reach from the interposer router to the boundary router of GPU-3.  

\section{Methodology}
\label{experimentation}
\subsection{Experimental Setup}

\begin{table}[t]
    \centering
    \resizebox{\columnwidth}{!}{
    \begin{tabular}{l|l}
        \hline
        Parameter & Value  \\ \hline\hline
        CPU &  2 GHz frequency, TimingSimple  \\ \hline
        \multirow{2}{*}{CPU Cache} & L1I and L1D - 32KB 4-way \\ 
               & L2 - 64KB 8-way \\ \hline
        GPU & 1 GHz frequency~\cite{amdgem5} \\ \hline
        \multirow{3}{*}{GPU Cache} & SQC (shared L1I) - 32KB, 8 way \\
                  & TCP (private L1D) - 16KB , 16 way\\
                  & TCC (Texture Cache per Channel) - 256 KB, 16 way \\ \hline
        Memory & Build-in memory model in Gem5~\cite{hansson2014simulating} \\ \hline
       \multirow{3}{*}{Network} & Booksim integraded with Gem5, 4-stage routers \\ 
                 & 1-flit buffers per control VC, 4-flit buffers per data VC \\
                 & 64 bit flit size and channel width \\ \hline    
    \end{tabular}
    }
    \caption{Parameters of simulated architecture.}
    \label{tab:architecture_parameter}
\end{table}

We use multiple simulation environments for performance evaluation, and hardware synthesis for design timing validation, area and power estimation. Gem5~\cite{binkert2011gem5} based full system setup is used for both simulating homogeneous (CPU-CPU) and heterogeneous (GPU-CPU) modular SoCs. We carry out detailed network deadlock study using synthetic traffic in BookSim~\cite{jiang2013detailed}. We model rc\_buffer, and {\it OPIC} system using Verilog HDL and synthesize using {\it TSMC 45nm} library along with router RTL~\cite{routerrtl}.

In full system setup, we integrate BookSim with Gem5 to simulate network of Compute Units (CUs) in the GPU chiplets (GCN-3~\cite{amdgem5}), CPUs in CPU chiplet, and also different chiplets on active interposer as summarized in Table~\ref{tab:architecture_parameter}.
We use 4 stage routers having 1-flit buffers per control VC and 4-flit buffers per data VC. Flit size and link channel width is 64 bit. The control packets are 1 flit and data packets are 5 flits. In homogeneous setup we configure SoC using multiple (${4\times4}$ mesh) CPU chiplets only and use MOESI hammer as the coherence protocol. Heterogeneous setup uses the multi-chiplet APU configuration~\cite{jiemingISCA2018}, consisting of four GPU chiplets (${4\times4}$ mesh, 16 CUs), one CPU chiplet (${2\times2}$), and an active interposer (${4\times4}$ mesh) as shown in Figure~\ref{fig:soc_network}. We use the in-built memory model in Gem5 equipped with eight memory channels and 8 banks per channel. 
 
We run heterogeneous system-level simulation on APU applications taken from AMD ROCm Developer Tools~\cite{HCCEx} and Rodinia~\cite{che2009rodinia} suites. We also evaluate {\it RC} in homogeneous full system setup using PARSEC~\cite{bienia2008parsec}. For thorough study of average packet latency with various configurations such as injection rates, VC/rc\_buffer sizes and chiplet sizes, we use synthetic traffic patterns in BookSim. Unless otherwise mentioned, for synthetic experiments packet size is 8 flits; we use four ${4\times4}$ chiplet and one ${2\times2}$ chiplet connected using ${4\times4}$ interposer network, having 2-VC-4-stage routers with 4-flit buffer depth and 4 packet space in the rc\_buffer. To be consistent with the full system setup, we also denote the chiplets as GPU and CPU chiplet in network-only simulations, where each node is treated as traffic source and sink. In addition, we can have any combination of routing algorithms in the chiplets. 

\subsection{Traffic Patterns}
In general, most of the communications happen among edge interposer routers and CPUs/GPUs. These interposer routers are attached with memory controller and coherence directory. Since CPU and GPU share memory space in GCN-3 configuration, large amounts of communications happen between CPU and coherence directory. Also, significant communications happen in between GPUs and directory controllers. {Communication with memory controller happens via directory for cacheable data.} There is low communication between GPU chiplets, and no direct communication between CPU and GPU chiplet. CPU and GPU communicate with each other using shared memory. For instance, in the current system setup, we can expect intra- and inter- chiplet, chiplet to interposer traffic including coherence and main memory requests and reply. Depending upon the applications' accesses, we may observe high similarity with {\it Uniform Random (UR)}, and {\it Bit-Complement (BC)} traffic patterns~\cite{jiemingISCA2018}. 
In addition, we extend the experiments with all the other synthetic traffic patterns, so that they can evaluate for other possible full system configurations.       

\section{Results}
\label{results}
In this section, we present the results and analysis for both synthetic and real workloads. For synthetic workloads we compare \textit{RC} with MTR, ITB and VC-SEP. For full system we compare performance of \textit{RC} with the state-of-the-art technique MTR.  

\subsection{Synthetic Workloads}
Figure~\ref{fig:throughput_graph} shows the load-latency curves for various traffic patterns. We first analyze the throughput and investigate reasons for throughput saturation in different techniques. Then we compare the low-load latency of these techniques across various traffic patterns.

\subsubsection{Throughput Analysis}
\label{throughput_analysis}
Figure~\ref{fig:throughput_graph} shows that \textit{RC} outperforms MTR, ITB, and VC-SEP in terms of network throughput in all the synthetic traffic patterns. We explain the throughput for UR as an representative of synthetic traffic patterns.  
\begin{figure}
    \centering
    \includegraphics[scale=0.31]{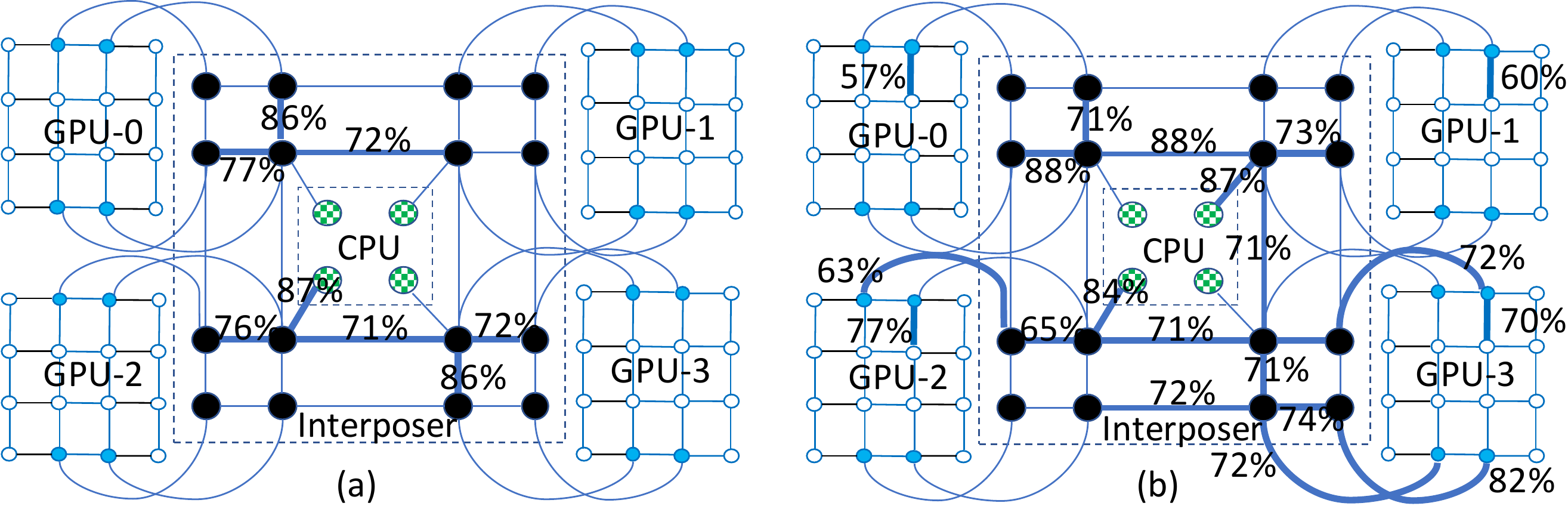}
    \caption{\footnotesize{Difference in maximum link utilization between (a) \textit{RC} and (b) MTR technique. Each number on the link represents the percentage of maximum number of cycles the link was busy across all the sample (10,000 cycles) periods.}}
    \label{fig:app_traffic_pattern}
\end{figure}
In {\it UR}, the source and destinations are generated randomly, where most traffics are inter-chiplet communication. 
For example, 3/4 of the generated traffics consists of outbound packets in the simulated configuration, which poses more stress on the boundary routers and interposer network. Across all the techniques, VC-SEP has least throughput, due to under utilization of buffer resources. While in MTR, turn restrictions on boundary routers create load imbalance, leading to throughput degradation. In ITB, ACK/NACKs packets are used for re-transmit request, data and control packets, which leads to higher network load and saturates the network earlier. In contrast, \textit{RC} regulates outbound packet injections facilitated by rc\_buffer in boundary routers and frees VC usage constraints for better resource utilization. Additionally, \textit{RC} provides routing flexibility so that traffics can be distributed evenly to all boundary routers. With these benefits, \textit{RC} improves throughput by up to 12\% and 117\% compared to the second best and the worst techniques, respectively.

To analyze the traffic distributions and communication bottlenecks of different designs, we depict hotspots as heatmap for UR for MTR, ITB, VC-SEP, and \textit{RC} at their near saturation load\footnote{Different techniques have different saturation load.} as shown in Figure~\ref{fig:hotspots}. Hotspot is defined as the average packet residency time in the router. Darker color represents higher packet residency time due to congestion. MTR imposes multiple extra turn-restrictions, resulting into hotspots due to imbalanced traffic distribution inside chiplets as shown in Figure~\ref{fig:modular_hotspots}, which leads to low network throughput as shown in Figure~\ref{fig:throughput_graph}. 
Heatmap for VC-SEP, as depicted in Figure~\ref{fig:vc_hotspots}, shows the severe congestion throughout the SoC network, which is due to the intensive usage of limited outbound VCs, making network saturation early. Interestingly, for ITB the contention inside the chiplets is very low, which can be attributed to packet drop~\cite{flich2003applying} that yields the buffer resources in the network. However, extra packet transmissions cause high energy and power consumption in the chiplets. Since \textit{RC} has uniform flow of packets as shown in Figure~\ref{fig:rc_hotspots}, it exhibits a better throughput than MTR and VC-SEP. \textit{RC} alleviates the long waiting of outbound packets from the chiplet routers. However, the contention in the interposer network partially offsets the throughput benefit, expected because of \textit{RC} scheme.
Note that we have plotted the heatmap with different injection rate (throughput injection rate) for each technique to show their distinct saturation behaviors, and point out the key reason for saturation. We notice that across all the techniques the interposer network is heavily used (outbound packets from multiple chiplet nodes go through one interposer router), which could be a bottleneck for achieving throughout improvements. 

\subsubsection{Latency Analysis}
As shown in Figure~\ref{fig:throughput_graph}, we observe a similar low load latency among \textit{RC}, MTR and VC-SEP across various traffic patterns. Figure~\ref{fig:zero_load_latency_break}a presents the detailed comparison of low-load latency for \textit{UR} as an example. It shows ITB increases a few more cycles as compared to other techniques. The unnecessary ejection and re-injection at low-load incurs two extra hops that causes high latency overhead. While in the interposer, MTR selects the minimum route to the destination node inside the chiplet, compensating the detours caused by turn restriction. Since \textit{RC} selects the minimum route to reach a chiplet instead of the destination node, it may incur more hops inside the chiplet. We expect to achieve better low-load latency for \textit{RC} if we incorporate more chiplet information by relaxing modularity constraints while designing interposer routing.

In Figure~\ref{fig:zero_load_latency_break}b, we show the average packet latency breakdown for \textit{RC} for UR to understand the overhead incurred by injection control. It shows that the portion of granting delay for rc\_buffer reservation over packet latency increases with the increase in network load at the beginning, and decreases while moving from medium load to high load. This is because at low load, rc\_buffer reservation causes constant delay without contention. Whereas at medium load, contention on rc\_buffers increases the granting delay. As injection rate increases to high load, the exponential injection queuing delay dominates the packet latency, which reduces the impact of rc\_buffer reservation significantly. In future, we plan to manage the rc\_buffers in a more proactive way, similar to token circulation rather than on-demand requesting to reduce the constant delay at low to medium load.

\subsection{Real Workloads}
\begin{figure}[t]
    \includegraphics[scale=0.28]{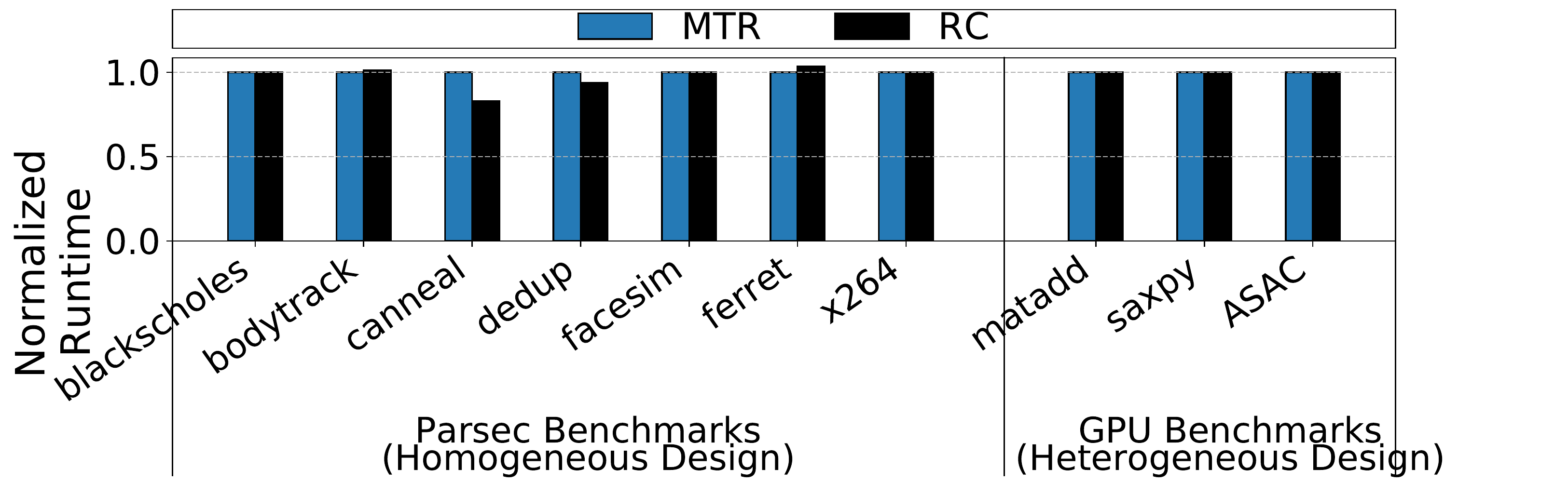}
    \caption{Normalized execution time full system simulations (lower is better). }
    \label{fig:full_system_exe_time}
\end{figure}

\subsubsection{Application Speedup}
Figure~\ref{fig:full_system_exe_time} shows the normalized full system execution time for homogeneous system evaluation using PARSEC benchmarks and heterogeneous system evaluation using GPU benchmarks, respectively. Since most PARSEC benchmarks have very low network load, \textit{RC} and MTR perform similarly for most of them. Among the benchmarks, \textit{RC} improves execution time over MTR for \textit{canneal} and \textit{dedup}. We observe that \textit{canneal} and \textit{dedup} have average injection rates of about 0.065 and 0.017 flit/node/cycle, respectively. It shows that \textit{canneal} has high netowrk load while \textit{dedup} has medium network load. According to phase behaviors~\cite{perelman2003using} during application execution, both benchmarks can saturate the network easily in their communication phase. Therefore, the better throughput provisioning of \textit{RC} improves the overall system performance for these two benchmarks. One outlier is \textit{ferret}, where \textit{RC} execute more instructions and experience more CPU idle time with higher variance. Our profiling analysis discovers that the extra instructions are spent on synchronization constructs which is implemented as spinning in \textit{ferret}. For the throughput oriented GPU benchmarks, the latency improvement is not directly translated to performance gain, where all the techniques achieve similar execution time.

\subsubsection{Link Utilization}
Figure~\ref{fig:app_traffic_pattern} projects the traffic distribution of \textit{Sync/AsyncAC (ASAC)} by showing the link utilization for both \textit{RC} and MTR. For both the techniques the interposer links are highly used for regular memory accesses as well as for offloading job from CPU to GPUs, and getting back results in the CPU through main-memory, connected with edge interposer routers. We observe that in MTR, outbound traffics is imbalanced towards the boundary routers due to turn restrictions, which is reflected in the uneven link utilization of boundary router links. In contrast, \textit{RC} distributes outbound traffics more evenly to the boundary routers, resulting in more network traffic balance.

\subsection{Routing Obliviousness}
\begin{figure}[t]
\centering
    \includegraphics[scale=0.25]{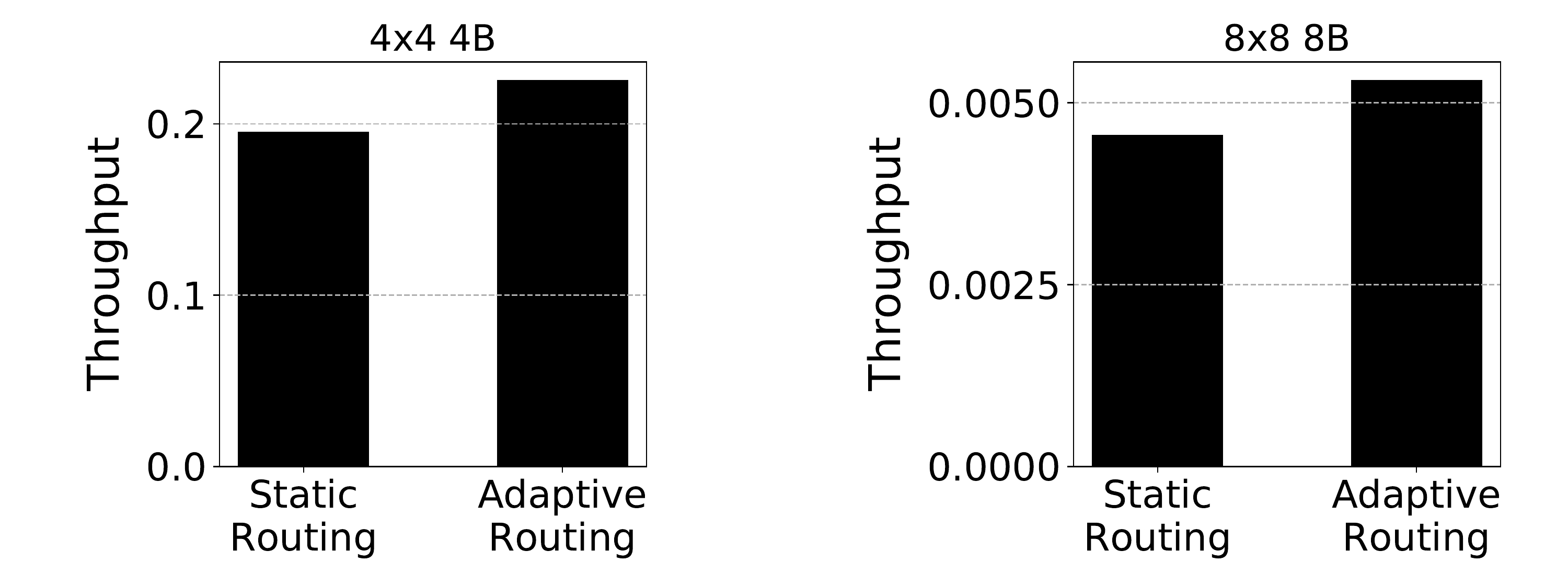}
    \caption{{\footnotesize{Throughput improvement by using adaptive routing for different size of modular SoCs for the following systems. The left two bars are for 4 boundaries in 4$\times$4 GPU chiplet and 4 boundaries in 2$\times$2 CPU chiplet. The right two bars 8 boundaries in 8$\times$8 GPU chiplet and 4 boundaries in 4$\times$4 CPU chiplet.}}}
    \label{routing_oblivious}
\end{figure}

In this section, we show \textit{RC} is routing oblivious by implementing Dynamic Credit-based Routing, where each router adaptively selects either XY, or YX routing depending on the credit availability in the downstream router. To demonstrate the benefits of routing oblivious \textit{RC}, we alleviate the bottleneck in interposer as discussed in Section~\ref{throughput_analysis} by providing 2 extra VCs only for interposer routers. In Figure~\ref{routing_oblivious}, we show that when dynamic routing is applied, the throughput improves in both smaller system (68 node, 84 routers) and bigger system (272 nodes, 304 routers) by 15.3\% and 21\%, respectively.  
\subsection{Sensitivity Analysis}
\label{analysis}
We scrutinize the system using various size and number of chiplets to obtain better understanding about system scalability with \textit{RC}. Difference of throughput is also observed with different VC sizes and increasing size of rc\_buffer. We intend to provide enough insight for estimating the best combination of these parameters for the system designers.   

\begin{figure}[t]
    \includegraphics[scale=0.28]{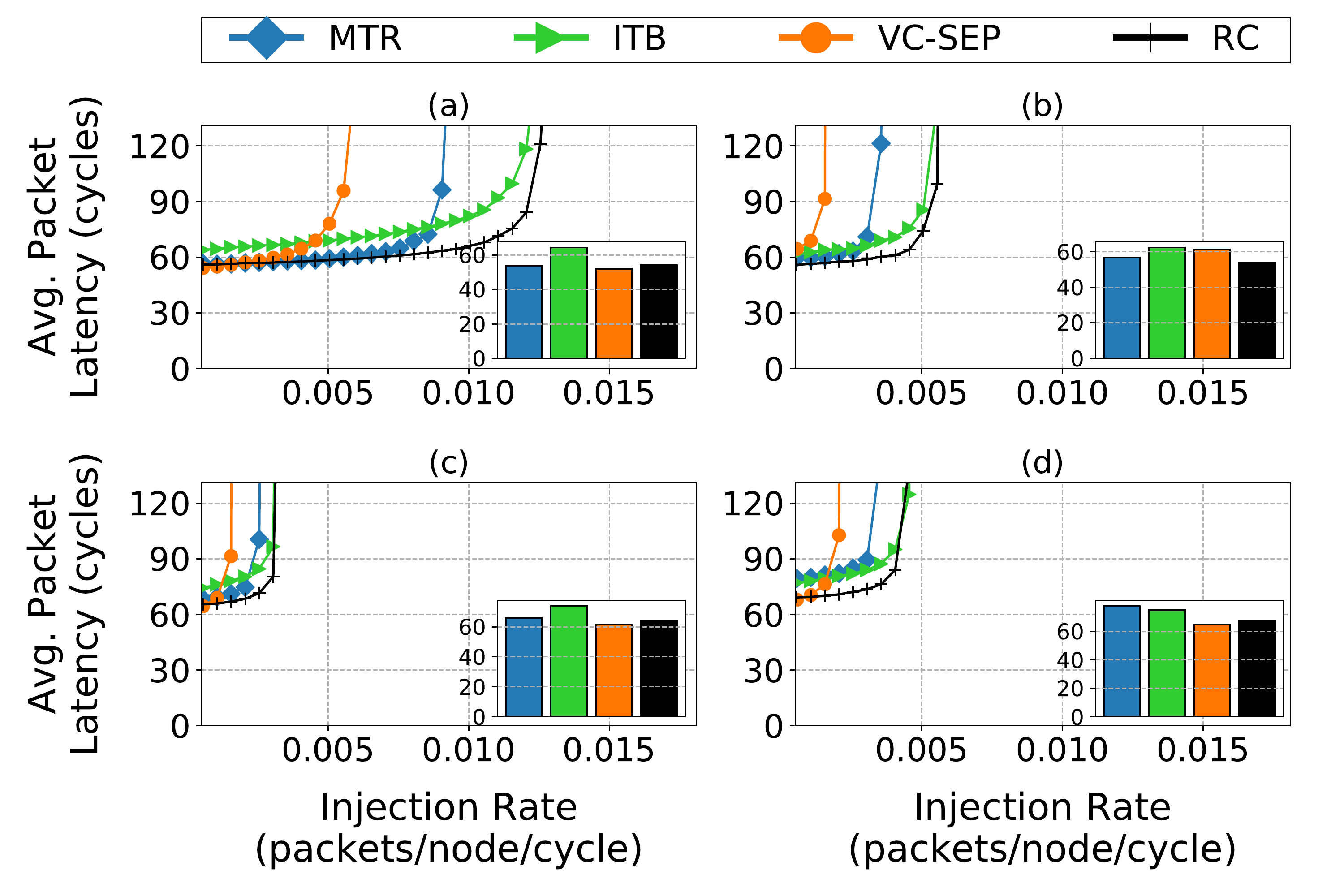}
    \caption{Sensitivity study: Scaling the chiplet size keeping the number of boundaries same as 4/chiplet. (a) Eight GPU chiplets of size 4x4 and one CPU chiplet of size 2x2 mesh. 
    (b) two GPU chiplets of size 8x8 and one CPU chiplet of size 2x2 mesh. (c) Four 8x8 GPU and one 4x4 CPU. (d) Same as (c) except 8 boundaries/GPU chiplet. The small bar chart in each of the graphs represents zero load latency for that particular configuration. }
    \label{parking_scalibility}
\end{figure}

\subsubsection{System Scalability}
We extensively study the system scalability as shown in Figure~\ref{parking_scalibility} by increasing number of chiplets in Figure~\ref{parking_scalibility}b (132 nodes) as compared to Figure~\ref{fig:throughput_graph} (68 nodes). To compare the scalability with different size of chiplets, we also keep the total number of nodes same between Figure~\ref{parking_scalibility}a (132 nodes) and Figure~\ref{parking_scalibility}b (132 nodes) and contrast their zero load latency and throughput. Figure~\ref{parking_scalibility}d shows a large system with large number of nodes per chiplet (total 272 nodes) with doubled number of boundaries in each chiplet. In all the configurations \textit{RC} outperforms MTR, ITB, and VC-SEP in terms of throughput. Also in terms of zero load latency \textit{RC} exhibits same or better than MTR and much better than ITB. This is because the detour caused by turn restrictions in MTR surpluses rc\_buffer request delay in \textit{RC}. For example, in Figure~\ref{parking_scalibility}c, MTR has 2 extra hops than \textit{RC}, which accounts for ~17\% more in average hops. We observe that the throughput difference reduces with the increase in the system size, as more nodes saturate the bisection bandwidth earlier. 

\begin{figure}[t]
    \centering
    \includegraphics[scale=0.27]{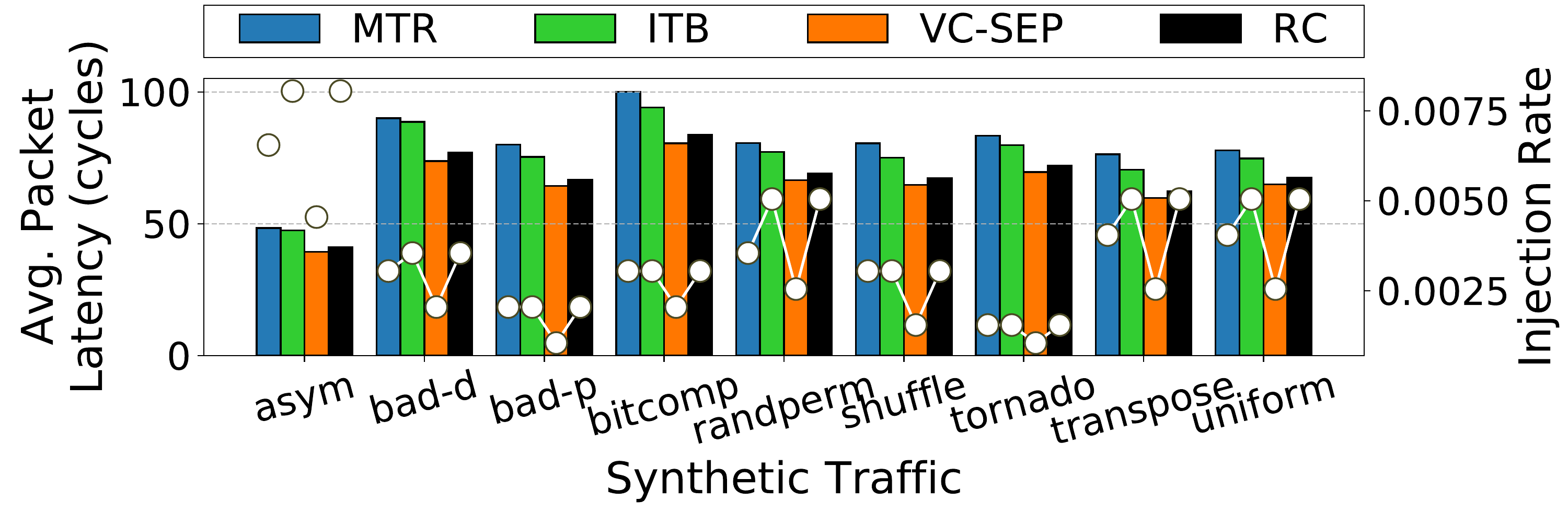}
    \caption{Doubled the number of boundaries 8 boundaries/ 8${\times}$8 GPU chiplet and 4 boundaries in 4${\times}$4 CPU chiplet. The major Y-axis corresponds to zero load latency shown in bar graphs, and minor Y-axis corresponds to the throughput as shown in white dots.}
    \label{parking_doubled_boundaries}
\end{figure}

\begin{figure}[t]
    \centering
    \includegraphics[scale=0.27]{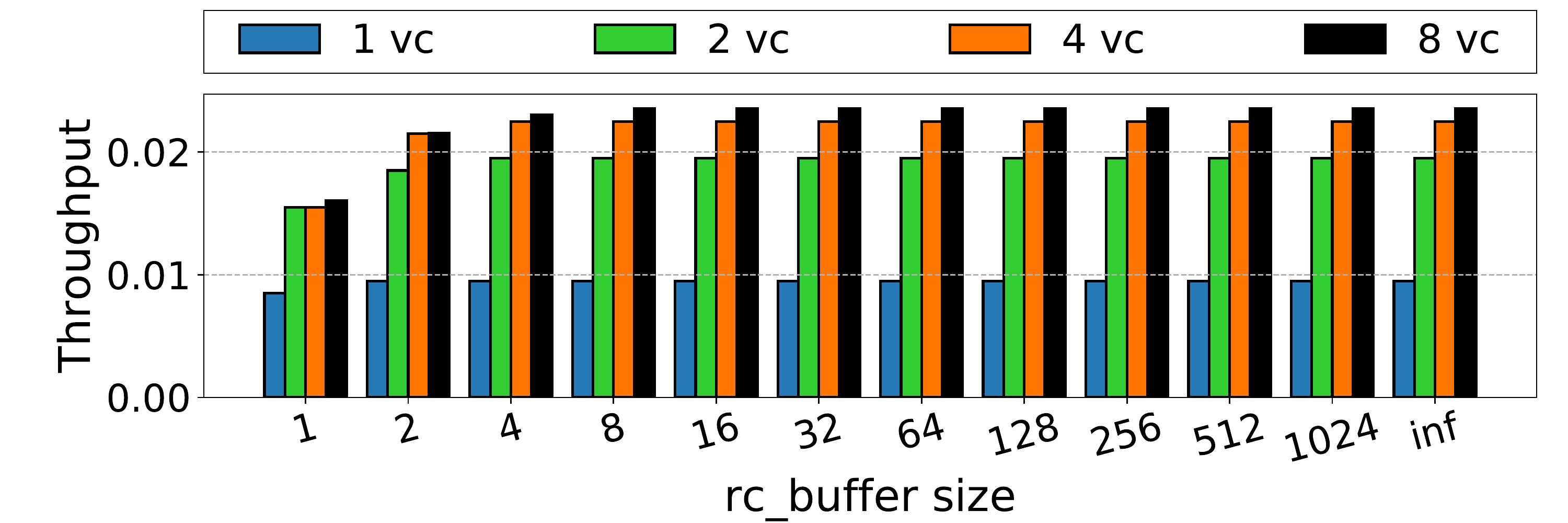}
    \caption{Throughput sensitivity and interplay between virtual channel and rc\_buffer size for 4$\times$4 chiplets (68 nodes setup) with 4 boundaries/chiplet.}
    \label{vc_and_parking_sizesensitivity}
\end{figure}

\begin{figure}[t]
    \centering
    \includegraphics[width=\columnwidth]{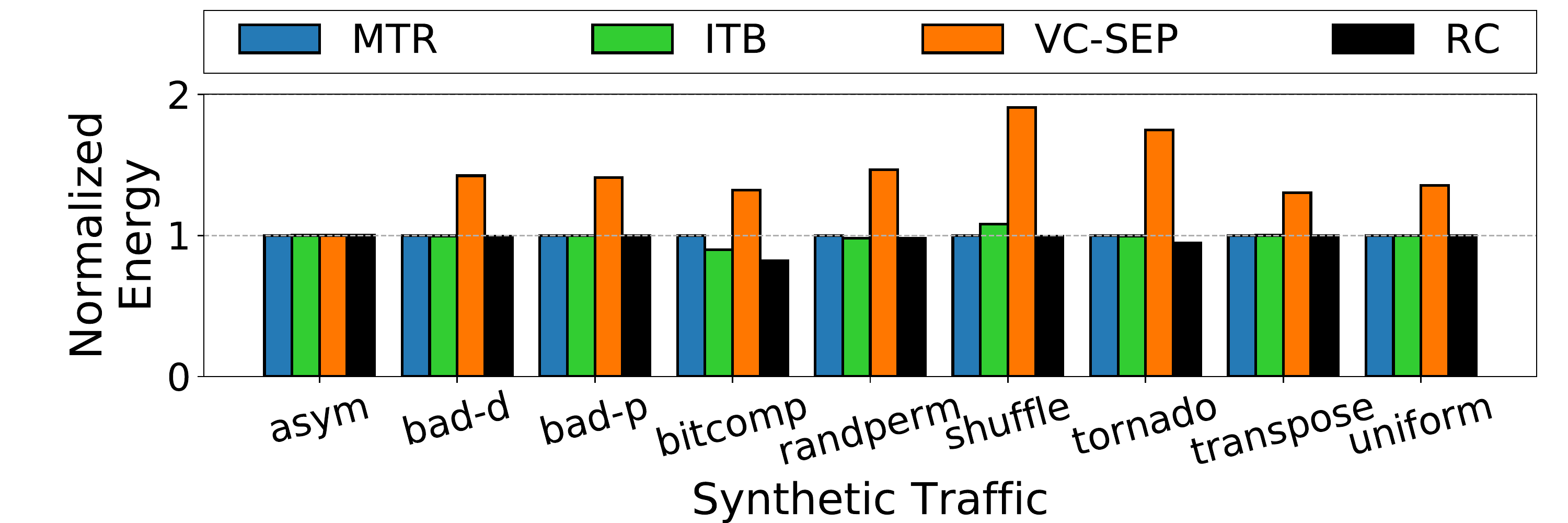}
    \caption{Normalized energy for all the techniques, across all the synthetic traffic patterns.}
    \label{parking_energy}
\end{figure}

We quantitatively show that with the increase in the chiplet size, overhead of OPIC does not hamper performance. In a ${4\times4}$ mesh with four boundaries, each boundary gets three requesters, and all of them get response from the boundary OPIC block in 2 cycle. However, in a ${8\times8}$ mesh network, maximum seven requester nodes can be connected as they are in one, or two hop distance from the boundary. In that case the furthest node from the boundary gets the response in total 6 cycles. Other nodes get response in much lesser time. Since the requests get registered in the next OPIC block, requester needs to send request only once. When we scale the number of nodes further we do not need to reconsider the setup time and hold time, as the amount of work need to be done in one cycle will still be same. Only the number of cycles of getting response will increase with the increase in distance from the boundary router.
However, it is worth noting that we opt for modular SoC design as we do not want to make large chips, rather want to put multiple small chiplets together to scale the system size. In Figure~\ref{parking_scalibility}a and in Figure~\ref{parking_scalibility}b, we show that system with multiple smaller chiplets in Figure~\ref{parking_scalibility}a has better throughput than system with fewer large chiplets. One reason for the difference in lower throughput with large number of smaller chiplets is the number of boundary routers are same for each chiplet both the systems~\cite{jiemingISCA2018}.   

Going one step further we doubled the number of boundary routers for 8$\times$8 chiplets keeping every other parameter values same. In Figure~\ref{parking_doubled_boundaries}, the result shows that the average packet latency in case of \textit{RC} improve significantly) over MTR by up to {17\%}. Figure~\ref{parking_scalibility}(c) and~\ref{parking_scalibility}(d) show results for SoCs with 4 boundaries and 8 boundaries per GPU chiplet, respectively. Average packet latency increased from 68 cycles to 80 cycles in case of MTR, and there is almost no change for our techniques. Our experimental results show that MTR travels more than 19.5\% extra hops as compared to \textit{RC}. This can be attributed to extra turn restrictions imposed by MTR. In addition, since the complexity of CDG analysis increased exponentially, we run the MTR algorithm for 7 days to explore the design space and pick the optimal result, which may not be the optimum turn restrictions for 8 boundary router setup. Interestingly, VC-SEP shows the zero load latency in this setup. However, the throughput suffer a lot because of low VC buffer utilization. In contrast, CDG analysis in 4 boundary setup takes only less than 2 hours to finish.

\subsubsection{Sensitivity to RC\_Buffer Size and VC Size}
In Figure~\ref{vc_and_parking_sizesensitivity}, we show impact of rc\_buffer size on network throughput, which is the saturation injection rate for SoC network with four 8$\times$8 GPU chiplets and one 4$\times$4 CPU chiplet (272 nodes), which shows similar trend for the smaller baseline setup with 4$\times$4 GPU chiplets.
We observe that increase of both rc\_buffer size and the number of VCs have impact on the system throughput. With rc\_buffer size of 1, we see hardly any throughput improvement with increasing VC sizes. The throughput improvement from single buffer to 2 buffer in rc\_buffer is almost 2$\times$. Also with 1-VC, increase in the \textit{RC} size improves throughput marginally. Result shows that for all the VC sizes, rc\_buffer size of 4 is good enough to provide achievable throughput, which is the case in infinite rc\_buffer, where the OPIC delay is zero. In addition, in terms of throughput the difference between 4-VC, and 8-VC result is also not very significant. Even 2-VC result also shows a good trade-off between throughput and energy consumption.  

\subsection{Area and Energy Analysis}

The hardware complexity and area overhead of \textit{RC} is very minimal. As per our detailed synthesis report, in each router of size 49667.53 ${\mu m^2}$, OPIC logic consumes only ${785.68 \mu m^2}$ area, which is 1.6\% of the router area. There are four rc\_buffer in each chiplet, and each has 4 packet buffers, consuming $6.0424 \mu m^2$ in total. Area overhead and hardware complexity incurred is negligible as compared to the total chiplet area and complexity.

Since we focus on the network deadlock aspect in this work, we estimate only the network energy to compare between MTR, ITB, VC-SEP and \textit{RC} using DSENT~\cite{dsent} and rc\_buffer access energy from RTL simulation. Figure~\ref{parking_energy} shows energy consumption of different techniques normalized to MTR under 0.013 packets/node/cycle injection rate for 100000 packets. It shows \textit{RC}, MTR and ITB consume similar energy across all the synthetic traffic patterns. In contrast, VC-SEP consumes more energy due to low utilization of VCs, leading to longer simulation time that consumes more static energy. We expect \textit{RC} to save more energy by reducing the static energy in high load since it sustains higher throughput.
\section{Related Works}
\label{related}
Deadlock avoidance mechanisms fan out in two distinct branches, namely VC and turn model based, and flow control based techniques. The first type either rely on turn restrictions, or on dedicated/ordered VC buffer for different traffic types/directions. On the other hand, flow control techniques either control the injection of packets, or ensures bubble in the buffer to avoid deadlocks. The state-of-the-art solves the new SoC deadlock issue using routing based turn restrictive technique while \textit{RC} follows flow control based deadlock avoidance. 

\subsection{VC and Turn Model Based}
Duato proposed escape-VC~\cite{duato1993new}, a theory for deadlock freedom for routing with cyclic channel dependency. 
Duato's theory can be applied for both deadlock avoidance~\cite{dally1993deadlock,gratz2008regional} and deadlock prevention~\cite{ma2011dbar,anjan1995efficient,duato2001general} techniques. Idea of escape channel cannot be applied directly in modular SoC as the packets in the escape-VC must be propagated using a deterministic deadlock free algorithm, which cannot be guaranteed in a modular SoC. 
Recently Ebrahimi et al.~\cite{ebrahimi2017ebda} propose {\it EbDa} that provides exclusive sets of VCs to isolate traffics (say, intra-chiplet traffic, and inter-chiplet, or outbound traffic) to avoid deadlocks. However, VC separation leads to lower utilization and is shown less attractive in MTR~\cite{jiemingISCA2018}, and we also find the same way.

Dally et al.~\cite{dally1988deadlock} propose to use two or more VCs in order to avoid the cyclic channel dependencies.
It ensures deadlock freedom by using total ordering of VCs. Even though this condition is sufficient to avoid the deadlock, it is not necessary~\cite{aniruddhISCA2018}. Extra VCs result in increase in the router area and energy consumption. 
Based on Dally's theory, a few other techniques have been proposed that use additional VCs~\cite{chen2016eyeriss,chiu2000odd,kim2008technology} to avoid deadlock. 
Another way to achieve strict order of reservation for the shared VCs is by imposing turn restrictions~\cite{isca1992_glass_turnmodel, aisopos2011ariadne, fu2011abacus} on the packet traversal.

\subsection{Flow Control Based}
For providing deadlock freedom, flow control techniques either regulate the injection~\cite{carrion1997flow,canwen2008dimensional} of the packets or allow a packet to go forward depending on the buffer occupancy~\cite{roscoe1987routing} in the ring. The second concept is coined as bubble flow control by Puente et al.~\cite{797388} and applied in torus network for the flow control in escape channel. This concept is being used in in-transit buffer for avoiding deadlock in k-ary n-cube torus network~\cite{ puente2001adaptive}, and extended later for irregular off-chip network~\cite{Flich:2000:PEN:335231.335235, flich2003applying}, worm-whole switching~\cite{chen2013worm}, torus cache-coherent NoCs~\cite{tc2015_ma_fbfc}. 

Recently Ramrakhani et al.~\cite{aniruddhISCA2018} propose {\it SPIN}, a synchronized flow control technique for deadlock prevention in flat network. 
It is very challenging to apply synchronized flow control in modular SoC, where the chiplets are designed independently, and connected through the interposer routers. Moreover, synchronization of packet movement among chiplets make the design very complicated.

\section{Conclusions}
\label{conclusions}
Chiplet-based system integration on an active interposer is a scalable and economic solution for improving system performance. As deadlock freedom is one of the main concerns, we propose \textit{RC}, a simple routing oblivious technique for modular SoCs. It completely protects the idea of modular design by providing total independence to the chiplet vendors, in terms of routing logic, topology, dimension, etc. The low load latency improvement of \textit{RC} over MTR, ITB and VC\_SEP are upto 15.49\%, 19.17\%, and 13.76\% across different configurations for all the synthetic workloads. The throughput improvement achieved by \textit{RC} over MTR, ITB, and VC-SEP are upto 56.34\%, 12.12\%, and 2.5$\times$, respectively.   
In full system simulation for real workloads, we marginally improve performance compared to the state-of-the-art. As part of the future work, we want to investigate application aware {\it OPIC} system, where critical packets can be prioritized in the rc\_buffer for a better system performance. 

\bibliographystyle{ieeetr}
\bibliography{ref}

\end{document}